\documentclass[lettersize,journal]{IEEEtran}

\usepackage{subfigure}
\usepackage{algorithm}
\usepackage{algpseudocode}
\usepackage{threeparttable}
\usepackage{adjustbox}
\usepackage{multirow}
\usepackage{colortbl}
\usepackage{arydshln}
\usepackage{enumitem} 
\usepackage{booktabs} 

\usepackage{amsmath,amsfonts}

\usepackage{array}
\usepackage[caption=false,font=normalsize,labelfont=sf,textfont=sf]{subfig}
\usepackage{textcomp}
\usepackage{stfloats}
\usepackage{verbatim}
\usepackage{graphicx}
\usepackage{url}
\usepackage{mdwlist}
\usepackage{balance}

\newcommand{\tabincell}[2]{\begin{tabular}{@{}#1@{}}#2\end{tabular}}

\def\BibTeX{{\rm B\kern-.05em{\sc i\kern-.025em b}\kern-.08em
    T\kern-.1667em\lower.7ex\hbox{E}\kern-.125emX}}

\begin{document}
\title{Yet another Improvement of Plantard Arithmetic for Faster Kyber on Low-end 32-bit IoT Devices}

\author{Junhao Huang, Haosong Zhao, Jipeng Zhang, Wangchen Dai, Lu Zhou\\ 
Ray C. C. Cheung,~\IEEEmembership{Senior Member,~IEEE,} \c{C}etin Kaya Ko\c{c},~\IEEEmembership{Fellow,~IEEE,} Donglong Chen
\IEEEcompsocitemizethanks{
  
  \IEEEcompsocthanksitem 
  Junhao Huang, Haosong Zhao and Donglong Chen are with Guangdong Provincial Key Laboratory of Interdisciplinary Research and Application for Data Science, BNU-HKBU United International College, Zhuhai, China. Junhao Huang and Haosong Zhao are also with Hong Kong Baptist University, Hong Kong, China. Email: \{huangjunhao, zhaohaosong, donglongchen\}@uic.edu.cn.  

  Jipeng Zhang and Lu Zhou are with Nanjing University of Aeronautics and Astronautics, Nanjing, China. E-mail: jp-zhang@outlook.com, lu.zhou@nuaa.edu.cn.
  
  \c{C}etin Kaya Ko\c{c} is with Nanjing University of Aeronautics and Astronautics (China), I$\check{\mathrm{g}}$d\i r University (Turkey), and University of California Santa Barbara (USA). E-mail: cetinkoc@ucsb.edu.

  Wangchen Dai and Donglong Chen are with Zhejiang Lab, Hangzhou, China. E-mail: w.dai@my.cityu.edu.hk.
  
  Ray C.C. Cheung is with City University of Hong Kong, Hong Kong, China. E-mail: r.cheung@cityu.edu.hk.

  Corresponding author: Donglong Chen.
  }
}

\markboth{IEEE TRANSACTIONS ON INFORMATION FORENSICS AND SECURITY}%
{How to use}

\maketitle

\begin{abstract}
    In 2022, the National Institute of Standards and Technology (NIST) made an announcement regarding the standardization of Post-Quantum Cryptography (PQC) candidates. Out of all the Key Encapsulation Mechanism (KEM) schemes, the CRYSTAL-Kyber emerged as the sole winner. This paper presents another improved version of Plantard arithmetic that could speed up Kyber implementations on two low-end 32-bit IoT platforms (ARM Cortex-M3 and RISC-V) without SIMD extensions. Specifically, we further enlarge the input range of the Plantard arithmetic without modifying its computation steps. After tailoring the Plantard arithmetic for Kyber's modulus, we show that the input range of the Plantard multiplication by a constant is at least $\mathbf{2.14\times}$ larger than the original design in TCHES2022. Then, two optimization techniques for efficient Plantard arithmetic on Cortex-M3 and RISC-V are presented. We show that the Plantard arithmetic supersedes both Montgomery and Barrett arithmetic on low-end 32-bit platforms. With the enlarged input range and the efficient implementation of the Plantard arithmetic on these platforms, we propose various optimization strategies for NTT/INTT. We minimize or entirely eliminate the modular reduction of coefficients in NTT/INTT by taking advantage of the larger input range of the proposed Plantard arithmetic on low-end 32-bit platforms. Furthermore, we propose two memory optimization strategies that reduce 23.50\%$\sim$28.31\% stack usage for the speed-version Kyber implementation when compared to its counterpart on Cortex-M4. The proposed optimizations make the speed-version implementation more feasible on low-end IoT devices. Thanks to the aforementioned optimizations, our NTT/INTT implementation shows considerable speedups compared to the state-of-the-art work. Overall, we demonstrate the applicability of the speed-version Kyber implementation on memory-constrained IoT platforms and set new speed records for Kyber on these platforms.
\end{abstract}

\section{Introduction}\label{sec:pre}

With the emergence of quantum computing technology, the existing Public Key Cryptographic (PKC) schemes face a significant threat. Shor's algorithm on quantum computers can solve the mathematical hard problems of these PKC schemes, such as the big number factorization problem of RSA, the Discrete Logarithm Problem (DLP) of ElGamal, and the Elliptic Curve Discrete Logarithm Problem (ECDLP) of Elliptic Curve Cryptography (ECC), in polynomial time. Although the creation of quantum computers that can crack these PKCs is still beyond realistic reach \cite{gidney2021factor}, the potential threat of quantum computers has prompted the cryptographic community to seek alternative solutions to replace the traditional PKC. In 2022, NIST announced four finalists to be standardized for its six-year Post-Quantum Cryptography (PQC) standardization project \cite{NIST:PQCStandard}. Among them, CRYSTAL-Kyber \cite{DBLP:conf/eurosp/BosDKLLSSSS18} is the only KEM standard, while CRYSTAL-Dilithium \cite{Dilithium}, FALCON \cite{prest2020falcon}, and SPHINCS$^+$ \cite{sphincs+} are three digital signature standards. Therefore, the efficient implementation of Kyber, the only PQC KEM standard, will have great significance for the future deployment of Kyber in realistic practices, such as cloud, edge, and Internet of Things (IoT).

The application of IoT has spread out in various fields, such as smart healthcare, home, and transportation domains. Increasingly, IoT devices have been utilized to gather sensitive data from individuals, corporations, military organizations, and governments. It is estimated that by 2030, there will be around 30 billion interconnected IoT devices \cite{Statista:IoT_predict}. With the rapid development of quantum computing, safeguarding sensitive data on numerous IoT devices against quantum computers is a pressing concern in the near future. Therefore, deploying NIST KEM standard Kyber for these IoT devices is essential.
The ARM Cortex-M4 platform is a NIST-recommended platform for evaluating PQC performance and memory consumption in IoT scenarios \cite{PQM4}. It has sufficient memory and a powerful Single Instruction Multiple Data (SIMD) extension for PQC implementation. However, there are also many low-end platforms, such as the 32-bit ARM Cortex-M3/M0 and RISC-V, which have more restricted power, computational resources, and memory compared to ARM Cortex-M4. The above restrictions on these platforms result in different implementation requirements for Kyber. Both memory consumption and efficiency are critical metrics for evaluating the applicability of Kyber on low-end IoT devices. Therefore, it is crucial to conduct more in-depth research to achieve efficient and feasible Kyber implementation on these IoT devices in the upcoming years.



Kyber \cite{DBLP:conf/eurosp/BosDKLLSSSS18} is a lattice-based cryptographic (LBC) scheme that relies on the theoretical security of the Module Learning-with-Error (MLWE) problem. Previous software implementations \cite{NewHope, seiler2018faster, botros2019memory, alkim2020cortex, DBLP:journals/tches/GreconiciKS21, abdulrahman2022faster} mainly focus on the improvement of LBC's core operations such as polynomial multiplication, vector inner product, and matrix-vector product.
Kyber's parameters enable efficient polynomial multiplication by using the Number Theoretic Transform (NTT), which reduces the time complexity of multiplying two degree-$n$ polynomials from $O(n^2)$ (using the schoolbook method) down to $O(n\log n)$. 
NTT's primary component is the butterfly unit, including two commonly-used algorithms: the Cooley-Turkey (CT) algorithm \cite{cooley1965algorithm} and the Gentleman-Sande (GS) algorithm \cite{gentleman1966fast}. A key operation in butterfly unit is the modular multiplication by a twiddle factor, where one operand is an arbitrary value, while the other one is a twiddle factor. Commonly, in software implementations, the twiddle factor is normally precomputed and stored in memory, hence the modular multiplication by the twiddle factor is treated as the modular multiplication by a constant.

Previous implementations \cite{NewHope,seiler2018faster,botros2019memory,alkim2020cortex,DBLP:journals/tches/GreconiciKS21,abdulrahman2022faster} utilized the Montgomery \cite{montgomery1985modular} or Barrett \cite{barrett1986implementing} arithmetic to compute the modular multiplication by the twiddle factor. However, neither of these methods fully utilized the fact that the twiddle factor is a precomputed constant to further speed up the operation. In 2021, Thomas Plantard \cite{Plantard:2021:Mod} proposed a word size modular arithmetic (Plantard arithmetic) that enables efficient computation of modular multiplication by a constant. 
Initially, the Plantard arithmetic only supported unsigned integers, which limited its feasibility in LBC schemes. 
Later in 2022, Huang et al. \cite{huang2022improved} presented an improved Plantard arithmetic tailored for the moduli in LBC schemes, which supports signed integers and has superior performance compared to the Montgomery and Barrett arithmetic. They achieved new speed records for Kyber and NTTRU by replacing the Montgomery and Barrett arithmetic with their signed-version Plantard arithmetic on Cortex-M4. However, their method is mainly benefited from utilizing the Cortex-M4's SIMD instruction \textbf{smulw\{b,t\}} to perform the \(16\times 32\)-bit multiplication and obtain the middle 16-bit effective product. The actual performance of the Plantard arithmetic remains unexplored on low-end 32-bit platforms without such a powerful SIMD extension. Moreover, the Plantard multiplication by a constant in \cite[Algorithm 11]{huang2022improved} supports inputs that are larger than a 16-bit signed integer, but the NTT/INTT implementation on Cortex-M4 limits the coefficients to 16-bit signed integers due to SIMD extension. Therefore, further investigation is necessary to determine whether the Plantard arithmetic can also replace the Montgomery and Barrett arithmetic on low-end 32-bit platforms without SIMD extension and whether the large input range can be fully utilized in NTT/INTT implementations.




\textbf{Contributions.} This paper aims to improve the signed-version Plantard arithmetic introduced in TCHES2022 \cite{huang2022improved} and explore its applicability in Kyber on three low-end 32-bit platforms: Cortex-M3 and two RISC-V platforms (SiFive Freedom E310 and PQRISCV). The contributions of this paper are as follows:
\begin{enumerate}
	\item We further enlarge the input range of the Plantard arithmetic without modifying its computation steps, and give the proof. After tailoring the Plantard arithmetic for Kyber's modulus, we show that the input range of the Plantard multiplication by a constant can be at least $2.14\times$ larger than the work in TCHES2022 \cite{huang2022improved}.
	\item We propose two optimization techniques to implement the Plantard arithmetic utilizing the specific Instruction Set Architecture (ISA) characteristics of Cortex-M3 and RISC-V. The optimized implementation shows that the Plantard arithmetic can indeed supersede the state-of-the-art Montgomery and Barrett arithmetic on these low-end 32-bit microprocessors.
	\item Based on the optimized Plantard arithmetic, we propose an efficient CT algorithm that consumes one instruction fewer than the GS algorithm on Cortex-M3. The optimized CT algorithm is then used to speed up the INTT implementation on this platform. 
	We apply 3-layer and 4-layer merging strategies to the NTT/INTT implementation on Cortex-M3 and RISC-V, respectively. 
	We show that the Plantard arithmetic with an enlarged input range enables a better lazy reduction strategy, minimizing or entirely eliminating the modular reduction of coefficients in NTT/INTT. Our NTT and INTT implementations achieve significant speedups compared with state-of-the-art, with NTT/INTT achieving speedups of 26.19\%/32.57\%, 34.76\%/55.53\%, and 22.67\%/43.37\% on Cortex-M3, SiFive RISC-V board, and PQRISCV, respectively. 
	\item We provide two versions of Kyber implementations, namely the stack-friendly (stack-version) and high-speed (speed-version) implementations on both Cortex-M3 and RISC-V, in line with \cite{abdulrahman2022faster}. For the speed-version implementation, we propose two memory optimization strategies that result in a significant reduction in stack usage, ranging between 
	23.50\%$\sim$28.31\% compared to that of \cite{abdulrahman2022faster}, making it more feasible on memory-constrained IoT devices. Overall, our optimized Kyber implementations achieve new speed records on both Cortex-M3 and RISC-V. Specifically, the speed-version outperforms PQM3 by a margin of 3.69\%$\sim$5.63\%, 3.51\%$\sim$5.15\%, and 3.37\%$\sim$4.67\% for Kyber512, Kyber768, and Kyber1024, respectively, on Cortex-M3. On PQRISCV, it outperforms previous work by 13.59\%$\sim$27.03\%, 25.49\%$\sim$31.15\%, and 26.96\%$\sim$31.43\% using only 26.86\%$\sim$52.44\% of their stack usage for the three Kyber variants, respectively.
\end{enumerate}

The remainder of this paper is organized as follows. In Section \ref{sec:background}, we review Kyber and its most time-consuming operations. In Section \ref{sec:modular_arith}, we show how to enlarge the input range of the Plantard arithmetic and tailor it for Kyber's modulus. The optimized implementations of the Plantard arithmetic as well as Kyber on Cortex-M3 and RISC-V are presented in Section \ref{sec:plantard_impl} and \ref{sec:apply_LBC}. We present and compare the experimental results in Section \ref{sec:results}. Finally, we conclude this paper in Section \ref{sec:conclusions}. 

\paragraph*{Availability of Our Software} The source codes of this paper is publicly available at \url{https://github.com/UIC-ESLAS/Kyber_RV_M3}.

\section{Preliminaries}\label{sec:background}
This section gives a brief introduction to Kyber, its underlying time-consuming operations and the target platforms.

\subsection{Kyber}
Kyber \cite{DBLP:conf/eurosp/BosDKLLSSSS18} is the only KEM standard in the NIST PQC standardization project. 
The overall security of the scheme is based on the theoretical security of the Module Learning With Errors (Module-LWE) problem, which was first introduced by Langlois and Stehl\'{e} in \cite{DBLP:journals/dcc/LangloisS15}. 
Specifically, for \((\mathbf{A}, \mathbf{b}=\mathbf{A}^T \mathbf{s}+\mathbf{e})\), the decisional Module-LWE problem describes the computational difficulty of distinguishing \((\mathbf{A}, \mathbf{b})\) from a uniform random pair, where \(\mathbf{s} \text{ and } \mathbf{e}\) denotes the secret and noise vectors sampled from a centered binomial distribution \(B_{\eta}(R_q^{k})\); \(\mathbf{A}\) is a public matrix sampled from a uniform random distribution \(\mathcal{U} (R_{q}^{k \times k})\). The small \(k\)-dimensional matrix and vector are introduced by the Module-LWE problem, which helps to balance the inefficiency of LWE \cite{DBLP:conf/stoc/Regev05} and the potential weakness of the structured Ring-LWE \cite{DBLP:conf/eurocrypt/LyubashevskyPR10}. \(k\) equals 2, 3, and 4 for the three Kyber variants, respectively.

The underlying polynomial ring of Kyber is \({R}_q=\mathbb{Z}_q[X]/(X^n+1)\), where \(q=3329\) and \(n=256\). The IND-CCA2-secure KEM of Kyber is constructed over an IND-CPA public-key encryption (PKC) scheme through the Fujisaki-Okamoto (FO) transform \cite{DBLP:conf/crypto/FujisakiO99}. We refer interested readers to \cite{avanzi2020crystals} for details of the PKC and KEM protocols of Kyber. Due to the introduction of the $k$-dimensional matrix and vector, the matrix-vector multiplication and vector inner product are two essential and time-consuming operations in Kyber. 

\subsection{Number Theoretic Transform}\label{subsec:ntt}
The underlying polynomial ring and its parameter choice ($q\text{ and } n$) of Kyber enable efficient polynomial multiplication with number theoretic transform (NTT). While using NTT, the polynomial \(f(X)\) in the polynomial ring  \(\mathbb{Z}_q[X]/f(X)\) can be factored as \(
    f(X)=\prod_{i=0}^{n-1} f_{i}(X)(\operatorname{mod} q)
\), where \(f_{i}(X)\) are small-degree polynomials. To multiply two degree-\((n-1)\) polynomials \(a, b \in \mathbb{Z}_q[X]/f(X)\) using NTT, the first step is to compute \(\hat{a}_i = a \operatorname{mod} f_i(X), \hat{b}_i = b \operatorname{mod} f_i(X) \), where \(i=0,\cdots,n-1\). Then, the polynomial multiplication of two degree-$n$ polynomials is divided into $n$ pairs of small-degree pointwise multiplications: \(\hat{a}_0\hat{b}_0,\cdots,\hat{a}_{n-1}\hat{b}_{n-1}\). Lastly, the inverse NTT (INTT) is used to get the result polynomial \(c=\text{INTT}(\text{NTT\textbf{}}(a)\circ \text{NTT}(b))\) in the normal domain. 

For Kyber, suppose that \(\zeta\) is the primitive \(256\)-th root of unity, the polynomial \(f(X)=X^{256}+1\) can be factored as
\( X^{256}+1=\prod_{i=0}^{127}\left(X^{2}-\zeta^{2 i+1}\right).\) Here, the NTT of Kyber is a 7-layer incomplete-NTT \cite{avanzi2020crystals} since \(f(X)\) is factored as degree-2 polynomials instead of degree-1.

NTT's primary component is the butterfly unit, including two commonly-used algorithms: CT algorithm~\cite{cooley1965algorithm} and GS algorithm~\cite{gentleman1966fast}. CT algorithm accepts normal order input but produces bit-reverse order output. In contrast, the GS algorithm takes bit-reverse order input and generates normal order output. To avoid the bit-reversal operation, the CT algorithm is normally used in NTT while the GS algorithm is adopted in INTT. However, recent work \cite{abdulrahman2022faster,AbdulrahmanCCHK22} shows that better performance could be achieved when CT algorithm is used in INTT on Cortex-M4.

\subsection{Modular Arithmetic}
\paragraph*{Notations} Before moving into details of the modular arithmetic, we first describe some notations used in the remaining sections.
We set \(l=16\) throughout the paper to support the 16-bit NTT/INTT in Kyber. 
We divide the modular reduction into $\bmod$ and $\operatorname{mod}^{\pm}$. For an integer $c$, $c \bmod q$ generates the modulo result of $c$ in $[0,q)$, while $c \operatorname{mod}^{\pm} q$ produces output in $[-\frac{q+1}{2},\frac{q}{2})$. For simplicity, \([X]_{l^{\prime}}\) denotes \((X \operatorname{mod}^{\pm} 2^{l^{\prime}})\), and \([X]^{l^{\prime}}\) means \((X >>l^{\prime})\) for a positive integer $l^{\prime}$.

Based on whether there is a constant operand, we divide the modular multiplication into two categories: modular multiplication by a constant and modular multiplication of two variables.
The multiplication inside the butterfly unit in NTT/INTT is the former case, while the latter one is used in the base multiplication of Kyber. Previous software implementations \cite{NewHope,seiler2018faster,botros2019memory,alkim2020cortex,DBLP:journals/tches/GreconiciKS21,abdulrahman2022faster} utilized the Montgomery multiplication \cite{montgomery1985modular} to perform these modular multiplications. The Barrett reduction \cite{barrett1986implementing} is normally used to reduce the range of the coefficients in NTT/INTT \cite{botros2019memory,alkim2020cortex,DBLP:journals/tches/GreconiciKS21}. However, the Montgomery arithmetic in previous software implementations offered the same implementation for two kinds of modular multiplications. In 2021, Thomas Plantard proposed the Plantard arithmetic, which offers an efficient modular multiplication by a constant. Later in 2022, Huang et al. \cite{huang2022improved} proposed a signed-version Plantard arithmetic and demonstrated its applicability in LBC schemes on Cortex-M4.

\begin{algorithm}[thb]
    \caption{Signed Montgomery multiplication \cite{seiler2018faster} \label{alg:signed_mont_rdc}}
    \begin{algorithmic}[1]
        \Require{Operand $a,b$ such that \(-q2^{l-1} \leq  a b<q2^{l-1}\), where $l$ is the machine word size, the odd modulus $q\in (0,2^{l-1})$}
        \Ensure{\(r \equiv a b 2^{-l} \bmod q, r\in (-q,q)\)}
        \State{\(c=c_{1} 2^l+c_{0}=a\cdot b \)}
        \State{\(m = [c_{0} \cdot q^{-1}]_{2l}\)}\Comment{   \(q^{-1}\) is a precomputed constant}
        \State{\(t_{1} =[m \cdot q]^l\)}\Comment{shift right operation}
        \State{\(r = c_{1}-t_{1}\)}
        \State{\(\)}\Return{\(r\)}
    \end{algorithmic}
\end{algorithm}

In this section, we only introduce two modular arithmetic, the Montgomery and Plantard arithmetic, that share some similarities. For details about the Barrett arithmetic, we refer to \cite{barrett1986implementing} and its signed implementation in Kyber\footnote{Signed Barrett reduction in Kyber's implementation \url{https://github.com/pq-crystals/kyber/blob/master/ref/reduce.c}}.
Algorithm \ref{alg:signed_mont_rdc} and \ref{alg:signed_plant} give brief descriptions for the signed-version Montgomery and Plantard multiplication, respectively. Both of them require three multiplications and some additions or shift operations to complete a modular multiplication. The outputs of them are not straightforward least positive modular results as in Barrett arithmetic. As suggested in \cite{NewHope}, when Montgomery-like algorithms are used in the modular multiplication by the twiddle factor in NTT, we can store the twiddle factor in the Montgomery or Plantard domain, namely multiplying the twiddle factor with $2^l$ or $2^{2l}$ mod $q$. In this way, the Montgomery/Plantard multiplication by the twiddle factor could cancel the special term ($2^{-l} \text{ or } 2^{-2l}$) and generate output in the normal domain.



\begin{algorithm}[t]
    \caption{Signed Plantard multiplication \cite{huang2022improved}}\label{alg:signed_plant}
    \begin{algorithmic}[1]
        \Require{Two signed integers $a,b\in [-q2^{\alpha},q2^{\alpha}],q <2^{l-\alpha-1}, q^{\prime}=q^{-1} \operatorname{mod}^{\pm} 2^{2l}$}
        \Ensure{$r= ab(-2^{-2l}) \operatorname{mod}^{\pm} q \text{ where }r\in [-\frac{q+1}{2},\frac{q}{2})$}
        \State{$r = \left[\left(\left[[abq^{\prime}]_{2 l}\right]^{l}+2^{\alpha}\right)q\right]^{l}$}
		\State{\Return{$r$}}
    \end{algorithmic}
\end{algorithm}

The biggest difference between these two algorithms is that the product $ab$ is only used once in the Plantard multiplication, while it is used twice in the Montgomery multiplication. Then, when operand $b$ is a constant, the Plantard multiplication can precompute the product of $b$ and $q^{-1}$ mod $2^{2l}$. 
Therefore, if we can handle the $l\times 2l$-bit multiplication $a\cdot (bq^{\prime})$ in one instruction on the target platform, the Plantard multiplication is one multiplication fewer than the Montgomery and Barrett arithmetic. 
Apart from this advantage, Huang et al. \cite{huang2022improved} further improved the Plantard arithmetic's input range and output range, making it a better modular arithmetic in LBC schemes.

\subsection{Target Platforms: Cortex-M3 and RISC-V}\label{subsec:platforms}


The ARM Cortex-M3 platform we use is Arduino Due which integrates an Atmel SAM3X8E core \cite{due2017arduino}. It comprises 96 KiB of RAM and 512 KiB of Flash, and operates at a maximum frequency of 16 MHz. Similar to Cortex-M4, the Cortex-M3 platform has 16 32-bit general-purpose registers, of which fourteen are programmable. However, it doesn't provide any floating-point (FP) registers and SIMD extensions. It should be noted that there are some non-constant-time instructions like \textbf{umull, smull, umlal} and \textbf{smlal} \cite{de2015performance} that we need to avoid using during the cryptographic implementation, otherwise it may suffer from timing attacks. The only two ``secure" multiplication instructions one can use in this platform are the \textbf{mul} and \textbf{mla} instructions for computing the low 32 bits of the 64-bit product. The \textbf{mul} instruction costs one cycle while the \textbf{mla} instruction takes two cycles. Most of the other instructions are 1-cycle, while the memory access operations such as \textbf{ldrh, ldr, ldrd, strh, str} and \textbf{strd} are relatively expensive. Loading or storing half-word/word requires two cycles, and loading or storing double-word (\textbf{ldrd/strd}) takes three cycles. Cortex-M3 also provides a useful inline barrel shifter operation, similar to Cortex-M4, which performs an additional shift operation before the addition and subtraction instructions, without additional overhead. For instance, the instruction (\textbf{add.w rd, rn, rm, asr \#16}) first right-shifts the operand \textbf{rm} by 16 bits, and subsequently, the addition is performed using the right-shifted result.

RISC-V is an open-source standard Instruction Set Architecture (ISA), enabling a new era of processor innovation through open collaboration. Because of its open-source features, it has attracted huge attention from both research community and industry. We select two RISC-V platforms as the benchmark platforms. The first one is the SiFive Freedom E310 board, which integrates a 32-bit E31 RISC-V core \cite{SiFive:Manual}. The RV32IMAC instruction set is instantiated in this core. This platform is a memory-constrained IoT device with only 16KiB of RAM, making it a suitable platform for measuring the performance and applicability of LBC schemes on low-end IoT devices. The second platform is a RISC-V simulator based on VexRiscv\footnote{\url{https://github.com/SpinalHDL/VexRiscv}}, as described in pqriscv-vexriscv\footnote{\url{https://github.com/mupq/pqriscv-vexriscv}} and used in PQRISC\footnote{\url{https://github.com/mupq/pqriscv}}. The RV32IM instruction set is instantiated in this simulator. In the following sections, we will denote the VexRiscv RISC-V simulator as PQRISCV. Both platforms comprise of 32 32-bit general-purpose registers, of which thirty are programmable. Similar to the Cortex-M3 platform, they do not have powerful SIMD extensions. The cycle counts of some instructions of PQRISCV are not well-specified in the documents, so we will not discuss them in detail. The cycle counts mentioned below only refer to the SiFive board. There are only two multiplication instructions (\textbf{mul, mulh}) that can be used, and both of them take five cycles on the SiFive board. Loading or storing one word (\textbf{lw, sw}) requires two cycles while loading or storing one half-word or byte (\textbf{lh, lhu, lb, lu}) takes three cycles. Apart from the multiplication, division, and memory access operations, most of the other instructions consume one cycle. All instructions are constant-time except the division.

\section{Improvements on Plantard Arithmetic}\label{sec:modular_arith}
In this section, the detail of the Plantard arithmetic with enlarged input range will be introduced. And this improved version is further tailored for Kyber's modulus. 

\subsection{Plantard Arithmetic with Enlarged Input Range}\label{subsec:improved_arith}
%

We observe that the signed-version Plantard multiplication (see Algorithm \ref{alg:signed_plant}) proposed in \cite{huang2022improved} accepts inputs $a,b$ in range $[-q2^{\alpha},q 2^{\alpha}]$. By replacing $q<2^{l-\alpha-1}$ into the range of $a,b$, we can see that it approximately covers most of the $l$-bit signed integers $(-2^{l-1},2^{l-1})$. Therefore, the Plantard reduction, which takes $a\cdot b$ as input, can only accept input in $(-2^{2l-2},2^{2l-2})$, and it cannot cover all the $2l$-bit signed integers (i.e., $[-2^{2l-1},2^{2l-1})$). 
As mentioned before, the coefficients in the NTT/INTT implementation in \cite{huang2022improved} cannot overflow 16-bit signed integer due to the use of the SIMD extension on Cortex-M4. Therefore, the signed-version Plantard arithmetic proposed in \cite{huang2022improved} is already quite sufficient for the implementation on Cortex-M4. However, their implementation did not fully utilize the large input range, which is larger than the maximum bound of a 16-bit signed integer, of the Plantard multiplication by a constant \cite[Algorithm 11]{huang2022improved}. 
On the contrary, the implementation on low-end 32-bit platforms stores each coefficient in a 32-bit register. It enables us to temporarily overflow 16-bit signed integer and fully utilize the large input range of the Plantard arithmetic. 

\newcommand{\plantredrange}{[q2^l-q2^{l+\alpha},2^{2l}-q2^{l+\alpha})}
\newcommand{\plantoutputrange}{[-\frac{q+1}{2},\frac{q}{2})}
Through carefully theoretical analysis, we found that one can maximize the input range of the Plantard arithmetic and tailor it for a specific modulus, which contributes to a better NTT/INTT implementation on low-end 32-bit platforms. The proposed Plantard multiplication with enlarged input range is shown in Algorithm \ref{alg:plant_extend}. The proposed algorithm does not modify any computation steps compared with Algorithm \ref{alg:signed_plant}, except that we manage to further enlarge the input range of $ab$ from $[-q^2 2^{2\alpha},q^2 2^{2\alpha}]$ to $\plantredrange$. Specifically, by replacing $q$ by $q<2^{l-\alpha-1}$ in the above input ranges, one could get a at least two times larger input range compared to the original design. 

\begin{algorithm}[t]
  \caption{Plantard multiplication with enlarged input range}
  \label{alg:plant_extend}
    \begin{algorithmic}[1]
        \Require{Two signed integers $a,b$ such that $ab\in \plantredrange, q<2^{l-\alpha-1}, q^{\prime}=q^{-1} \operatorname{mod}^{\pm} 2^{2l}$}
        \Ensure{$r= ab(-2^{-2l}) \operatorname{mod}^{\pm} q \text{ where }r\in \plantoutputrange$}
        \State{$r = \left[\left(\left[[abq^{\prime}]_{2 l}\right]^{l}+2^{\alpha}\right)q\right]^{l}$}
	    \State{\Return{$r$}}
    \end{algorithmic}
\end{algorithm}

\subsection{Theoretical Proof}\label{subsec:proof}

Note that in 2024, Yang et al. \cite{yang2024modular} identified a misuse of the floor and ceiling functions in the correctness proof of improved Plantard arithmetic in \cite{huang2022improved}. The authors of the improved Plantard arithmetic then updated their correctness proof to rectify this issue in their eprint version \cite{cryptoeprint:2022/956}. The main update they made to the algorithm was to ensure $\alpha>0$ instead of $\alpha\geq 0$ in the original paper. Mostly, the theoretical proof of the Plantard multiplication with enlarged input range is similar to the one given in \cite{Plantard:2021:Mod}, \cite{huang2022improved} and \cite{cryptoeprint:2022/956}. Theorem \ref{the:correct} shows the correctness of Algorithm~\ref{alg:plant_extend}.

\newtheorem{theorem}{\bf Theorem}
\begin{theorem}[Correctness]{Let $q$ be an odd modulus, $l$ be the minimum word length (i.e., power of 2 number) 
    such that $q<2^{l-\alpha-1}$, where $\alpha> 0$, then Algorithm \ref{alg:plant_extend} is correct for $ab \in \plantredrange$.}
 \label{the:correct}
\end{theorem}

\renewcommand{\IEEEQED}{\IEEEQEDopen}

\begin{IEEEproof}[Proof of Theorem \ref{the:correct}] All preconditions in Algorithm \ref{alg:plant_extend} remain the same as in Algorithm \ref{alg:signed_plant} except the range of $ab$. We show that the input range of the Plantard multiplication can be maximized without modifying any computation steps. 
Similar to the proof in \cite{huang2022improved}, the correctness of this algorithm depends on the following four conditions:
  \begin{enumerate}[label=(\roman*)]
      \item $r\in \plantoutputrange$;
      \item $pq-ab$ is divisible by $2^{2l}$, where $p\equiv abq^{-1}\operatorname{mod} 2^{2l}$;
      \item $ 0<q 2^{l+\alpha}-p_{0} q+ab<{2^{2 l}}$, where $p_1=\left\lfloor \frac{p}{2^l}\right\rfloor, p_0=p-p_1 2^l$. Because the right-shift of the signed integer always rounds the result towards negative infinity, we always have $p_0\in[0,2^{l})$.
      \item $r= ab(-2^{-2l})\operatorname{mod}^{\pm} q$.
  \end{enumerate}
  
Since we do not modify any preconditions and computation steps of the Plantard multiplication, the first two conditions remain true. More specifically, for condition $r\in \plantoutputrange$, since $\left\lfloor\frac{abq^{\prime} \operatorname{mod}^{\pm} 2^{2l}}{2^l}\right\rfloor $ is the only part containing $ab$ and it is still in $[-2^{l-1},2^{l-1}-1]$ even though the range of $ab$ is increased. So, $r$ is still in $\plantoutputrange$.
For condition that $pq-ab$ is divisible by $2^{2l}$, because $q$ is an odd modulus, there always exist a $p\equiv abq^{-1}\operatorname{mod} 2^{2l}$ such that $pq-ab \equiv  0 \operatorname{mod} 2^{2 l}$. So, this condition is also valid.

The third condition is the most important part that demonstrates the correctness of the Plantard multiplication. 
In order to maximize its input range, we first assume that this condition holds for a new range of $ab$, i.e.,
\begin{equation}
    0< q 2^{l+\alpha}-p_{0} q+ab<{2^{2 l}},
    \label{eq:inequality}
\end{equation} 
where $q<2^{l-\alpha-1}, p= abq^{-1}\operatorname{mod}^{\pm} 2^{2l},p_1=\left\lfloor \frac{p}{2^l}\right\rfloor, p_0=p-p_1 2^l, p_0 \in [0,2^l)$ and $\alpha > 0$. Then, we can deduce the maximum input range of $ab$ through Inequality (\ref{eq:inequality}) as follows.

\begin{itemize}

\item When $ab \geq 0$ and $ 0 \leq p_0 < 2^l$, to ensure that the Inequality (\ref{eq:inequality}) holds, for the left-hand side of the inequality, we have:
\begin{gather*}
q 2^{l+\alpha}-p_{0} q+ab \geq q2^{l+\alpha}-p_{0} q > q2^{l+\alpha}-q2^{l}> 0,
\end{gather*}
which always holds for $\alpha > 0$. As for the right-hand side, we have:
\begin{gather*}
  q 2^{l+\alpha}-p_{0} q+ab \leq q 2^{l+\alpha}+ab <2^{2l}.
\end{gather*}
To ensure that the Inequality (\ref{eq:inequality}) holds, we conclude that $ab<2^{2l}-q2^{l+\alpha}$.

\item When $ab < 0$ and $ 0 \leq p_0 < 2^l$, for the right-hand side, we have:
\begin{gather*}
q 2^{l+\alpha}-p_{0} q+ab< q2^{l+\alpha}<2^{l-\alpha-1}2^{l+\alpha}<2^{2l},
\end{gather*}
which always holds for $\alpha > 0$. As for the left-hand side, we have:
\begin{gather*}
q 2^{l+\alpha}-p_{0} q+ab > q2^{l+\alpha}-q2^l+ab >0.
\end{gather*}

\end{itemize}

To ensure that the Inequality (\ref{eq:inequality}) holds, we conclude that $ab \geq q2^l-q2^{l+\alpha}$.
Therefore, the Inequality (\ref{eq:inequality}) is valid for every $ab\in \plantredrange$. 
By dividing every components in Inequality (\ref{eq:inequality}) by $2^{2l}$, we can get
\begin{equation*}
0<\frac{q2^{l+\alpha}-p_0q+ab}{2^{2l}}<1.
\end{equation*}
Overall, we conclude that 
\begin{align*}
r & = ab\left(-2^{-2 l}\right) \operatorname{mod} q \equiv \frac{pq-ab}{2^{2 l}}\\
&\equiv \left\lfloor\frac{pq-ab}{2^{2l}}+\frac{q2^{l+\alpha}-p_0q+ab}{2^{2l}}\right\rfloor\\
&\equiv \left\lfloor \frac{q\left(\frac{abq^{-1}\operatorname{mod}^{\pm} 2^{2l}}{2^l} +2^{\alpha}\right)}{2^{l}}\right\rfloor\\
\end{align*}

Finally, we obtain that $r=ab(-2^{-2l})\operatorname{mod}^{\pm} q =\left[\left(\left[[abq^{\prime}]_{2l}\right]^{l}+2^{\alpha}\right)q\right]^{l} $ for signed inputs $a$ and $b$ that satisfy $ab\in \plantredrange$.

\end{IEEEproof}

\subsection{Plantard Multiplication by a Constant Tailored for Kyber}\label{subsec:plant_kyber}

As mentioned before, Plantard arithmetic enables efficient modular multiplication by a constant in NTT/INTT, where the constant is the precomputed twiddle factor. During precomputation, one could always make sure that the twiddle factors are reduced to $[0,q)$. As the product $ab$ lies in $\plantredrange$, when we limit the constant $b$ in $[0,q)$, the input range of $a$ can be further enlarged.

For Kyber, we have $q=3329$, $l=16$. To ensure that modulus restriction $q<2^{l-\alpha-1}$ holds, the maximum $\alpha$ is equal to 3. Besides, the maximum value of $b$ (i.e. $b_{max}$) is equal to $3328$ in Kyber. Overall, we can deduce the maximum and minimum value of $a$ by:
\begin{gather*}
    a_{max} < ({2^{2 l}-q2^{l+\alpha}})/b_{max}\\
    =(2^{32}-3329\times 2^{19})/3328\approx 230.13q.
\end{gather*}
\begin{gather*}
    a_{min}> (q2^l-q 2^{l+\alpha})/b_{max}\\
    =(3329\times 2^{16}-3329\times 2^{19})/3328\approx -137.85q.
\end{gather*}
\newcommand{\plantrange}{[-137q,230q]}

Thanks to our improvements, the modular multiplication by a constant in Kyber supports input of range $a\in \plantrange$. Compared to the range $[-64q,64q]$ in \cite[Algorithm 11]{huang2022improved}, the input range of the Plantard multiplication by a constant tailored for Kyber is at least $2.14\times$ larger.
\section{Efficient Plantard Arithmetic for 16-bit modulus on Cortex-M3 and RISC-V}\label{sec:plantard_impl}

The efficient Plantard arithmetic on Cortex-M4 \cite{huang2022improved} was dependent on the SIMD instruction \textbf{smulw\{b,t\}} for carrying out the $16\times 32$-bit multiplication $[a\times bq^{\prime}]_{2l}$. Here, $a$ denotes a $16$-bit signed integer and $bq^{\prime}$ represents a precomputed $32$-bit positive integer. In \cite[Algorithm 16]{huang2022improved}, Huang et al. also presented a 5-instruction implementation of the Plantard multiplication by a constant on RISC-V. They concluded that in cases where multiplication instruction is slower than the shift instruction, the Plantard arithmetic has better performance than the Montgomery and Barrett arithmetic. In this section, we propose two optimization techniques to improve the performance of the Plantard arithmetic based on the specific ISA characteristics of Cortex-M3 and RISC-V, respectively. These optimization techniques show that the Plantard arithmetic can outperform the Montgomery and Barrett arithmetic without the prerequisite of slow multiplication. This is demonstrated by the efficient implementation of Plantard arithmetic on Cortex-M3, where both multiplication and shift instructions take one cycle. Notably, the efficient implementation of Plantard arithmetic is not restricted to Kyber's modulus and can be customized for other 16-bit odd moduli as well.

\subsection{Plantard multiplication by a constant}
The first optimization technique is proposed for efficient Plantard multiplication by a constant on Cortex-M3 (Algorithm \ref{alg:mod_mul_M3}). Unlike the Plantard multiplication by a constant on Cortex-M4 \cite[Algorithm 11]{huang2022improved}, this algorithm accepts a 32-bit signed integer as input on low-end 32-bit platforms. If one implements the multiplication $[a\times bq^{\prime}]_{2l}$ with the $32\times 32$-bit multiplication instruction \textbf{mul}, the valid product locates in the most significant half of the 32-bit register $r$. Then, one need to shift the valid product to the least positive half so as to compute the later steps. The above shift operation might decrease the efficiency of Plantard arithmetic, as it consumes an extra cycle. In order to improve the efficiency, we propose to merge this shift operation with the ``add to $2^\alpha$" step by utilizing the inline barrel shifter operation on Cortex-M3 (cf. line 3 of Algorithm \ref{alg:mod_mul_M3}). This operation initiates by right-shifting register $r$ by 16 bits. Thereafter, the addition of $2^\alpha$ is directly carried out in the same cycle. After that, the remaining operations are computed by using just one multiplication and one right-shift instructions. In Section \ref{subsubsec:butterfly}, a similar technique will be proposed to omit the final shift operation in the CT algorithm on Cortex-M3.

\begin{algorithm}[tb]
  \caption{Efficient Plantard multiplication by a constant for Kyber on Cortex-M3}\label{alg:mod_mul_M3}
  \begin{algorithmic}[1]
        \Require{An $32$-bit signed integer $a\in \plantrange$, a precomputed $32$-bit integer $bq^{\prime}$ where $b$ is a constant and $q^{\prime}=q^{-1} \operatorname{mod}^{\pm} 2^{32}$}
        \Ensure{$r= ab(-2^{-2l})\operatorname{mod}^{\pm} q$}
    \State{$bq^{\prime} \leftarrow bq^{-1} \operatorname{mod} 2^{2l}$\Comment{precomputed}}
    \State{\textbf{mul} $r, a, bq^{\prime}$ }\Comment{$r\leftarrow [abq^{\prime}]_{2l}$}
    \State{\textbf{add} $r, 2^\alpha, r, \text{asr}\#16$}\Comment{$r\leftarrow ([r]^l+2^{\alpha})$}
    \State{\textbf{mul} $r, r, q$}
    \State{\textbf{asr} $r,r,\#16$}\Comment{$r\leftarrow [rq]^l$}
    \State{\Return{$r$}}
    \end{algorithmic}
\end{algorithm}

The second optimization technique for the Plantard multiplication by a constant is carried out on RISC-V, as demonstrated in Algorithm \ref{alg:mod_mul_risc_v}. Similar to the algorithm on Cortex-M3, we also utilize the $32\times 32$-bit multiplication instruction \textbf{mul} to implement the multiplication $[a\times bq^{\prime}]_{2l}$. As RISC-V does not provide inline barrel shifter operation as in Cortex-M3, we perform the right-shift operation and the addition of $2^{\alpha}$ separately. Then, instead of computing $[rq]^l$ with one multiplication and one shift similar to the final two steps in Algorithm \ref{alg:mod_mul_M3}, we propose to precompute $q2^l=q\times 2^{l}$, which is inspired by the similar technique of Montgomery arithmetic proposed in \cite{greconici2020kyber}. This technique utilizes the fact that $[rq]^l$ is equivalent to $[rq2^l]^{2l}$, and the division by $2^{2l}$ operation can be implemented by using one \textbf{mulh} instruction. Note that this optimization technique cannot be applied to Cortex-M3 as it does not offer constant-time full multiplication instruction, but it can be securely extended to other platforms as long as the targeted platform has constant-time \textbf{mulh} instruction.

For platforms that lack of inline shifter and \textbf{mulh} instructions, if there is an efficient and constant-time \textbf{mla} instruction to handle the multiply-and-add operation, it is also possible to merge step 4 and 5 in Algorithm \cite[Algorithm 16]{huang2022improved}. But we choose not to do it on Cortex-M3 since the \textbf{mla} instruction use two cycles on this platform, which will make this version slower than that in Algorithm \ref{alg:mod_mul_M3}.

\begin{algorithm}[tb]
    \caption{Efficient Plantard multiplication by a constant for Kyber on RISC-V}\label{alg:mod_mul_risc_v}
    \begin{algorithmic}[1]
          \Require{An $32$-bit signed integer $a\in \plantrange$, a precomputed $32$-bit integer $bq^{\prime}$ where $b$ is a constant and $q^{\prime}=q^{-1} \operatorname{mod} 2^{32}, q2^l=q\times 2^l$}
          \Ensure{$r= ab(-2^{-2l})\operatorname{mod}^{\pm} q$}
      \State{$bq^{\prime} \leftarrow bq^{-1} \operatorname{mod}^{\pm} 2^{2l}$\Comment{precomputed}}
      \State{\textbf{mul} $r, a, bq^{\prime}$ }\Comment{$r\leftarrow [abq^{\prime}]_{2l}$}
      \State{\textbf{srai} $r, r, \#16$}
      \State{\textbf{addi} $r, r, 2^\alpha$}\Comment{$r\leftarrow ([r]^l+2^{\alpha})$}
      \State{\textbf{mulh} $r, r, q2^{l}$}\Comment{$r\leftarrow [rq2^{l}]^{2l}$}
      \State{\Return{$r$}}
      \end{algorithmic}
\end{algorithm}

In summary, the efficient Plantard multiplication by a constant on both Cortex-M3 and RISC-V take four instructions, including two multiplications, one shift and one addition. Compared to the existing Plantard and Montgomery multiplication on RISC-V or Cortex-M3 \cite{huang2022improved,greconici2020kyber} and \cite{PQM3} that take five instructions, the proposed implementation is one multiplication fewer.

\subsection{Plantard reduction}\label{subsec:plant_red}
We propose a 4-instruction Plantard reduction for the modular multiplication of two variables on both architectures, which is equally efficient as the state-of-the-art Montgomery or Barrett reduction in \cite{greconici2020kyber} and \cite{PQM3}. The proposed instruction sequence of the Plantard reduction is similar to Algorithm \ref{alg:mod_mul_M3} and \ref{alg:mod_mul_risc_v}, except that the $32$-bit precomputed constant $bq^{\prime}$ is replaced by $q^{-1} \bmod 2^{2l}$ in the modular multiplication of two variables or $(-2^{2l} \bmod q)\times q^{-1}\bmod 2^{2l}$ in the modular reduction of coefficients. This variant of modular multiplication is used in the base multiplication of Kyber. Besides, previous implementations mostly utilize the Barrett reduction to implement the modular reduction of coefficients. Nevertheless, despite the same cycle consumption as the Barrett reduction on Cortex-M3 and RISC-V, we chose to adopt the Plantard reduction for the sake of consistency.

\section{Optimized Kyber Implementation on Cortex-M3 and RISC-V}\label{sec:apply_LBC}
Based on the efficient Plantard arithmetic presented in Section \ref{sec:plantard_impl}, we propose a number of optimizations that are geared towards improving the efficiency and reducing the memory consumption of Kyber on Cortex-M3 and RISC-V.

\subsection{Efficient 16-bit NTT/INTT Implementation}
This section presents efficient 16-bit NTT/INTT implementations on Cortex-M3 and RISC-V. 

\subsubsection{Butterfly Unit}\label{subsubsec:butterfly}
To integrate the Plantard arithmetic to the butterfly unit, we follow the method described in \cite{huang2022improved} to precompute the 32-bit twiddle factor as $\zeta =  ((\zeta \cdot (-2^{2l}) \operatorname{mod} q)\cdot q^{-1}) \operatorname{mod}^{\pm} 2^{2l}$. For two signed coefficients $a \text{ and } b$, the CT algorithm computes $a'=a+b \zeta \text{ and } b'= a-b \zeta$, while the GS algorithm computes $a'=a+b \text{ and } b'= (a-b)\cdot \zeta$. We present the efficient implementations of CT algorithm on Cortex-M3 and RISC-V in Algorithm \ref{alg:ct_butterfly_M3} and \ref{alg:ct_butterfly_RISC-V}, respectively. The instruction sequence for the GS algorithm is similar, hence we will not discuss it in detail.

We demonstrated that by utilizing the inline barrel shifter operation on Cortex-M3, one could omit the final shift operation in Algorithm \ref{alg:mod_mul_M3} and achieve an efficient 5-instruction CT algorithm with the Plantard arithmetic. Conversely, the GS algorithm cannot benefit from the barrel shifter operation due to the operation difference. Specifically, GS algorithm requires one more instruction compared to the CT algorithm. Nevertheless, it is worth highlighting that both CT and GS algorithms are one instruction fewer than their Montgomery-based counterparts on Cortex-M3 \cite{PQM3}. Since the CT algorithm costs one instruction fewer than the GS algorithm, it motivates us to also adopt it in INTT. 

Following similar optimization strategies outlined in \cite{abdulrahman2022faster} and \cite{AbdulrahmanCCHK22}, in order to minimize the side-effect of additional twists in the last layer of INTT using CT algorithm, light butterfly was proposed and applied in the first three layers of INTT. The so-called light butterfly is to reduce the modular multiplication by the twiddle factor in the CT butterfly, i.e., only computing $a'=a+b \text{ and }b'=a-b$. 
Additionally, they suggested that using CT algorithm in INTT could also help to decrease the number of modular reductions and yield a superior INTT implementation on Cortex-M4 \cite{abdulrahman2022faster}. After applying the CT algorithm to INTT of Kyber, we further validate its extensibility on Cortex-M3. It is worth to note that we are the first to apply the above improved CT algorithm to INTT on Cortex-M3. 

\begin{algorithm}[tb]
    \caption{CT algorithm on Cortex-M3}\label{alg:ct_butterfly_M3}
    \begin{algorithmic}[1]
    \Require{Two signed integers $a,b$, the 32-bit twiddle factor $\zeta$}
    \Ensure{$a'=a+b\zeta, b'= a-b\zeta$}
    \State{\textbf{mul} $b, b, \zeta$ }
    \State{\textbf{add} $b, 2^\alpha, b$, asr \#16}
    \State{\textbf{mul} $t, b, q$}
    \State{\textbf{sub} $b', a, t$, asr \#16}
    \State{\textbf{add} $a', a, t$, asr \#16}
    \State{\Return{$a',b'$}}
    \end{algorithmic}
\end{algorithm}

Algorithm \ref{alg:ct_butterfly_RISC-V} demonstrates the instruction sequence of the CT algorithm on RISC-V. Unlike the implementation on Cortex-M3, the GS and CT algorithms are equally efficient on RISC-V. Initially, we also contemplated the use of CT algorithm in INTT on RISC-V; however, the results indicated that the INTT with CT algorithm cannot outperform the GS one. The crucial reasons for that are two-fold. Firstly, the use of CT algorithm in INTT results in the doubling of the twiddle factors, which increases the memory access overhead. Specifically, in our implementation, the INTT with CT algorithm consumes 61 additional memory accesses for the twiddle factors, which reduces its efficiency and warrants a further 0.5KiB memory to accommodate these extra twiddle factors. The extra memory cost also increases the difficulty of deploying Kyber on memory-constrained platforms. 
Secondly, the number of modular reduction in INTT with GS algorithm has been significantly reduced by using Algorithm~\ref{alg:plant_extend} on RISC-V (see the better lazy reduction strategy in Section \ref{subsubsec:lazy_reduction} for detailed discussions). Thus, there left only a small space for further optimization through the utilization of CT algorithm in INTT.

\begin{algorithm}[tb]
    \caption{CT algorithm on RISC-V}\label{alg:ct_butterfly_RISC-V}
    \begin{algorithmic}[1]
    \Require{Two signed integers $a,b$, the 32-bit twiddle factor $\zeta$}
    \Ensure{$a'=a+b\zeta, b'= a-b\zeta$}
    \State{\textbf{mul} $b, b, \zeta$ }
    \State{\textbf{srai} $b, b$, \#16}
    \State{\textbf{addi} $b, b, 2^\alpha$}
    \State{\textbf{mulh} $t, b, q2^l$}
    \State{\textbf{sub} $b', a, t$}
    \State{\textbf{add} $a', a, t$}
    \State{\Return{$a',b'$}}
    \end{algorithmic}
\end{algorithm}

In summary, the optimal combination of butterfly usage on Cortex-M3 is to apply CT algorithm on both NTT and INTT. On the other hand, the most effective pattern on RISC-V is to use CT algorithm on NTT but GS algorithm on INTT.

\subsubsection{Layer Merging}\label{subsubsec:layer_merging}
Layer merging is an essential optimization strategy in NTT/INTT when memory access operations are costly. This strategy effectively reduces the expensive memory access by loading multiple coefficients at once and processing more NTT/INTT layers over these coefficients. The design of optimal layer merging strategy depends on the number of available registers on the target platform. 
For example, fourteen registers are programming-available on Cortex-M3. Thus, a maximum of eight coefficients can be loaded at once, allowing to merge three layers of NTT/INTT on this platform. The remaining six registers are utilized to store addresses or execute the modular arithmetic in the butterfly unit. On RISC-V, an abundance of 30 programming-available registers are provided with greater flexibility. Therefore, sixteen coefficients can be loaded at once, allowing to merge four layers of NTT/INTT at once, and there are still 14 registers left for other usage. Consequently, the 4-layer merging strategy is better on RISC-V.

In summary, for the 7-layer NTT/INTT in Kyber, we adopt the 3+3+1 and 3+1+3 layer merging strategy for NTT and INTT, respectively, on Cortex-M3. The motivation behind the choice of the 3+1+3 layer merging strategy instead of 3+3+1 will be discussed in Section \ref{subsubsec:lazy_reduction}. On the other hand, the 4+3 and 3+4 layer merging strategy are used for NTT and INTT on RISC-V, respectively.

\subsubsection{Better Lazy Reduction}\label{subsubsec:lazy_reduction}
The so-called ``lazy reduction strategy'' means deferring the modular reduction of coefficients until the value of coefficients exceed the maximum bound of the register or the input range of the subsequent operation. The input range of the modular multiplication by a constant in the butterfly unit is particularly crucial in facilitating this strategy. The NTT implementation with CT algorithm algorithm using Plantard arithmetic only increases the coefficients by $3.5q$. The large input range and small output range of Plantard arithmetic enables us to eliminate all the modular reductions of coefficients in NTT on Cortex-M3 and RISC-V. This is a noteworthy improvement compared to the Montgomery-based NTT implementation, which requires one modular reduction for all coefficients. Therefore, the focus of this study is on the exploration of better lazy reduction strategy for INTT on Cortex-M3 and RISC-V.

An even better lazy reduction strategy is designed for INTT in this work, compared to the version on Cortex-M4 in \cite{huang2022improved}. Two main factors contribute to this new design. Firstly, Huang et al. \cite{huang2022improved} utilized the SIMD extension of Cortex-M4 to speed up NTT/INTT; thus the coefficient in INTT cannot exceed the bound of a 16-bit signed integer. For Kyber's modulus $q=3329$, the coefficients must stay within $[-9.5q,9.5q]$. However, on the low-end 32-bit platforms that lack of SIMD extension, each coefficient is loaded to a 32-bit register, which has the potential to temporarily overflow 16-bit signed integer range before it is stored back to a 16-bit memory. Secondly, the input range of the Plantard arithmetic is enlarged from $[-64q,64q]$ in \cite[Algorithm 11]{huang2022improved} to $\plantrange$, which allows for a ``lazier" modular reduction.

\paragraph{Input Polynomial of INTT} Before moving to the detailed strategy, let's analyze the coefficient size of the input polynomial of INTT first. It is worth noting that matrix-vector multiplication and vector inner product are the two operations that could expand the coefficient size of INTT's input polynomial. The resulting polynomial of these two operations must add up $k$ intermediate polynomials, where $k$ equals $2,3,4$ for the three Kyber variants, respectively. Abdulrahman et al. \cite{abdulrahman2022faster} proposed two versions of implementations for these two operations and its underlying pointwise multiplication: the stack-version and the speed-version. In this study, we apply the Plantard arithmetic to both versions of implementations.

\begin{itemize}
    \item The stack-version process involves reducing each intermediate polynomial and accumulating $k$ reduced polynomials together, which generates an ``unreduced" polynomial (i.e., every coefficient of the polynomial is beyond the output range of modular reduction). When implemented with the Plantard arithmetic, the stack-version matrix-vector product and vector inner product produce coefficients within the range $(-kq/2,kq/2)$.
    \item The speed-version implementation is built on top of the stack-version implementation. An intermediate 32-bit array is used to cache the sum of $k$ intermediate polynomials, and this intermediate result is reduced after the accumulation process, producing a ``reduced" polynomial (i.e., every coefficient is in the output range of modular reduction). When implemented with the Plantard arithmetic, the speed-version code generates coefficients in the range $(-q/2,q/2)$.
\end{itemize}

\paragraph{Better Lazy Reduction on RISC-V}\label{para:better_lazy_RV} We first introduce a better lazy reduction strategy for INTT on RISC-V. Since INTT is computed by using GS algorithm in combination with the 3+4 layer merging strategy on RISC-V, after the third layer of INTT, the coefficients have to be stored back to 16-bit signed integers in memory. Therefore, we need to analyze the coefficient size to ensure that they can fit in 16-bit signed integer after the third layer of INTT.

\begin{figure}[t]
    \begin{small}
        \begin{center}
            \includegraphics[width=0.45\textwidth]{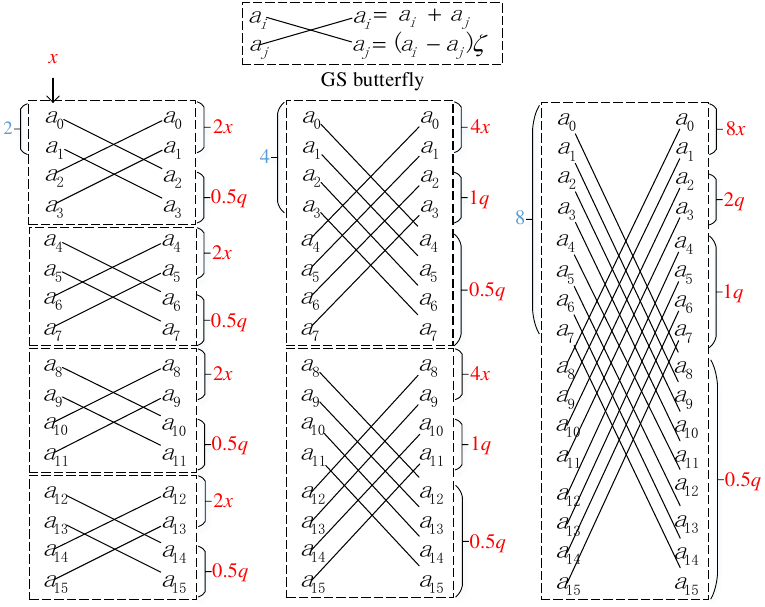}
        \end{center}
        \caption{Example of the first 16 coefficients of the first three layers of a length-128 INTT using GS algorithm on RISC-V. $a_i$ and $a_j$ represent coefficients of polynomial $a$. The red number $x$ in upper left corner represents the maximum absolute value of the input coefficient of INTT. Dashed rectangle represents GS butterflies of various step size; the computation details of GS butterfly are described in the topmost dashed rectangle. The blue number on the left-hand side of the rectangle indicates the step size of the butterfly unit, while the red number represents the maximum absolute value of the corresponding coefficients after the computation of each layer.}
        \label{fig:intt}
    \end{small}
\end{figure}

Fig. \ref{fig:intt} demonstrates the first 16 coefficients of the first three layers of a length-128 INTT implemented via GS algorithm. Since the input coefficients of the stack-version INTT are within the bound $(-kq/2,kq/2)$, we let the red number $x$ in Fig. \ref{fig:intt} equal to $kq/2$, which is the largest absolute value of the input range. After three layers of GS butterflies computation, the first 2 coefficients (out of 16 coefficients) are expanded up to $8x$, namely $4kq$. 
In the case of Kyber512 ($k=2$), the term $4kq$ is equal to $8q$, and it does not overflow 16-bit signed integer after the third layer. The later four layers of INTT produce coefficients that reach up to $128q$, which exceeds the input range of the Plantard arithmetic in \cite[Algorithm 11]{huang2022improved} $[-64q,64q]$, but remains within the input range of the proposed Plantard arithmetic with enlarged input range $\plantrange$. These coefficients will be reduced by the modular multiplication with the twiddle factors and $128^{-1}$ in the last layer of INTT. Hence, modular reduction can be completely avoided in the INTT of Kyber512.
As for Kyber768 and Kyber1024, where $k=3$ and $k=4$ respectively, the term $4kq$ is equal to $12q \text{ and } 16q$, respectively. Both $12q \text{ and } 16q$ overflow a 16-bit signed integer. Therefore, modular reductions are needed for these coefficients (2 out of 16). Due to the symmetric property of INTT, the entire INTT for Kyber768 and Kyber1024 requires modular reductions for $256\times 2/16=32$ coefficients only. By using the 3+4 layer merging strategy, the last four layers of INTT only expand some coefficients, e.g., $a_2$ and $a_3$ in Fig. \ref{fig:intt}, from $2q$ up to $32q$. These coefficients will also be reduced by the modular multiplication with the twiddle factors and $128^{-1}$ in the last layer of INTT. Therefore, no further modular reduction is needed.

The input coefficients of the speed-version INTT are within the bound $(-q/2,q/2)$ for all three Kyber variants. We let the red number $x$ in Fig. \ref{fig:intt} equal to $q/2$, which is the largest absolute value of the input range. After three layers' computation, the absolute value of these coefficients are smaller than $8x=4q$, which is well below the maximum value of a 16-bit signed integer and can be securely stored back to the 16-bit memory. After seven layers of INTT computation by using GS algorithm, these coefficients are expanded to $64q$. However, this is still within the input range of the proposed Plantard arithmetic. Thus, we are able to eliminate all the modular reduction of the speed-version INTT for the three Kyber variants.

\begin{figure}[t]
    \begin{small}
        \begin{center}
            \includegraphics[width=0.45\textwidth]{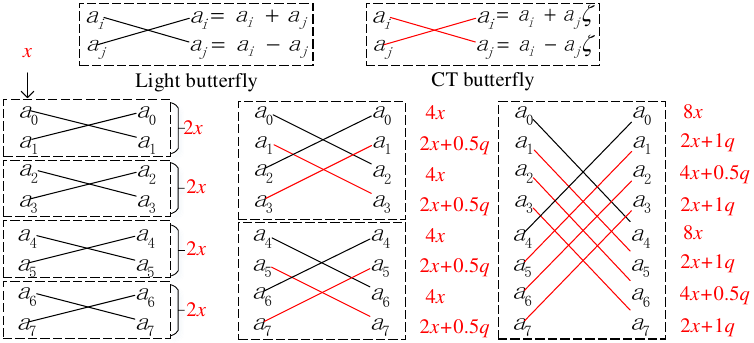}
        \end{center}
        \caption{Example of the first 8 coefficients of the first three layers of a length-128 INTT by using CT algorithm on Cortex-M3. $a_i$ and $a_j$ represent coefficients of the polynomial $a$. The red number $x$ in upper left corner represents the maximum absolute value of the input coefficient of INTT. Dashed rectangle represents light or CT butterflies of various step size; the computation details of light butterfly and CT algorithm are described in the topmost two dashed rectangles; the butterflies with black crossing line denote light butterfly while the butterflies with red crossing line represent CT algorithm; the red number represents the maximum absolute value of the corresponding coefficients after the computation of each layer.}
        \label{fig:intt_ct}
    \end{small}
\end{figure}

\paragraph{Better Lazy Reduction on Cortex-M3} We now describe a better lazy reduction strategy for INTT on Cortex-M3. The INTT on Cortex-M3 is computed by using CT algorithm in combination with the 3+1+3 layer merging strategy. After the third and fourth layer of INTT, the coefficients have to be stored back to 16-bit signed integers in memory. Therefore, we need to analyze the coefficient size and ensure that they can fit in 16-bit signed integer after the third and fourth layer.

Fig. \ref{fig:intt_ct} illustrates the first 8 coefficients of the first three layers of a length-128 INTT by using light butterfly and CT algorithm. As shown in the topmost dashed rectangles of Fig. \ref{fig:intt_ct}, the light butterfly proposed in \cite{abdulrahman2022faster} only involves one addition and one subtraction, omitting one modular multiplication by the twiddle factor of the CT algorithm. 
Consequently, each light butterfly will double the size of both coefficients while each CT butterfly will only increase each coefficient by $0.5q$.
For the stack-version INTT implementation, the absolute value of the input coefficients is smaller than $kq/2$. If we let the red number $x$ in Fig. \ref{fig:intt_ct} equal to $kq/2$, the computation of three layers INTT would expand $256\times 2/8=64$ coefficients (e.g., $a_0$ and $a_4$ in Fig. \ref{fig:intt_ct}) up to $8x=4kq$. Similar to the analysis in Section \ref{para:better_lazy_RV}, the term $4kq$ overflows a 16-bit signed integer for Kyber768 and Kyber1024. Therefore, one might need modular reductions for 64 coefficients after the third layer of INTT.
However, in order to minimize the number of modular reductions, we curtail half of the modular reductions by performing modular reductions for $256\times 1/8=32$ coefficients (e.g., $a_0$ in Fig. \ref{fig:intt_ct}) after the second layer of INTT.
After that, $a_0$ is reduced down to $0.5q$, and the third layer of INTT would only expand $a_0$ and $a_4$ in Fig. \ref{fig:intt_ct} up to $4x+0.5q$, which is equal to $6.5q/8.5q$ for Kyber768/Kyber1024. These numbers could fit in 16-bit signed integers. 
Since we use the 3+1+3 layer merging strategy and adopt the CT algorithm in the fourth layer, the fourth layer of INTT with CT algorithm would only increase the coefficients by $0.5q$. 
Therefore, the maximum absolute value of coefficients is $9q$ after the fourth layer, which can still fit in a 16-bit signed integer.
The reason why we adopt the 3+1+3 layer merging strategy instead of 3+3+1 is that it limits the growth of coefficients and obviates the need for more modular reductions. 
If we adopt the 3+3+1 layer merging strategy, the second three layers of INTT with CT algorithm would expand $256\times 4/8=128$ coefficients (e.g., $a_0,a_2,a_4$ and $a_6$ in Fig. \ref{fig:intt_ct}) up to $10q$, which will overflow a 16-bit signed integer, thus necessitating more modular reductions.

The final three layers of INTT in the 3+1+3 layer merging strategy increase the coefficients up from $9q$ to $72q$. It should be noted that $72q$ likewise surpasses the input range of $[-64q,64q]$ in Plantard arithmetic~\cite[Algorithm 11]{huang2022improved}, but can still fit into the input range $\plantrange$ of the proposed Plantard arithmetic. This further demonstrates the contribution of the larger input range to the better lazy reduction strategy. Similar to the INTT using GS algorithm on RISC-V, we also eliminate the modular reduction of coefficients for the stack-version INTT of Kyber512 and speed-version INTT of all three Kyber variants on Cortex-M3.

Overall, we only need modular reduction for 32 coefficients in the stack-version INTT of Kyber768 and Kyber1024 on both platforms. Conversely, we entirely eliminate the modular reduction of coefficients in the stack-version INTT of Kyber512 and the speed-version INTT of all three Kyber variants. The better lazy reduction strategy is made possible by leveraging the excellent merits of the Plantard arithmetic, the optimal layer merging strategies, and the 32-bit platforms' properties.

\subsection{Pointwise Multiplication and Memory Optimizations}\label{subsec:memory_opt}
\subsubsection{Pointwise multiplication}\label{subsubsec:basemul}
After the 7-layer NTT transform, two polynomials $a$ and $b$ are transformed into their NTT domain $\hat{a}$ and $\hat{b}$. 
The pointwise multiplication $\hat{c}=\hat{a}\circ \hat{b}$ is performed over $\mathbb{Z}_q[X]/(X^2-\zeta^{2\mathrm{br}_7(i)+1})$. The symbol $\mathrm{br}_7(i)$ denotes the bit reversal operation of a 7-bit integer $i\in[0,127]$ to obtain the corresponding twiddle factor. The symbol $\circ$ consists of 128 pointwise multiplications, and each of them is performed as $\hat{c}_{2i}+\hat{c}_{2i+1}X=(\hat{a}_{2 i}+\hat{a}_{2 i+1} X)(\hat{b}_{2 i}+\hat{b}_{2 i+1} X) \bmod (X^2-\zeta^{2 \mathrm{br}_7(i)+1})$, where $\hat{c}_{2 i}=\hat{a}_{2 i} \hat{b}_{2 i}+\hat{a}_{2 i+1} \hat{b}_{2 i+1} \zeta^{2 \mathrm{br}_7(i)+1}$ and $\hat{c}_{2 i+1}=\hat{a}_{2 i} \hat{b}_{2 i+1}+\hat{b}_{2 i} \hat{a}_{2 i+1}$.
As mentioned in Section \ref{subsubsec:lazy_reduction}, to align with the stack-version and speed-version matrix-vector multiplication, we also implement two versions of pointwise multiplications for Kyber. 

For the stack-version pointwise multiplication, we follow the lazy reduction strategy proposed in \cite{alkim2020cortex}, namely $\hat{c}_{2 i}=(\hat{a}_{2 i}\cdot \hat{b}_{2 i}+\hat{a}_{2 i+1}\cdot (\hat{b}_{2 i+1} \cdot \zeta^{2 \mathrm{br}_7(i)+1}\bmod q))\bmod q$ and $\hat{c}_{2 i+1}=(\hat{a}_{2 i}\cdot \hat{b}_{2 i+1}+\hat{b}_{2 i}\cdot \hat{a}_{2 i+1}) \bmod q$. In total, three modular reductions are required for each pointwise multiplication and the result is stored in 16-bit array. We can see that the computation of $\hat{b}_{2 i+1} \cdot \zeta^{2 \mathrm{br}_7(i)+1}\bmod q$ is a modular multiplication by a twiddle factor $\zeta^{2\mathrm{br}_7(i)+1}$ and can be speeded up with the efficient Plantard multiplication by a constant (Algorithm \ref{alg:mod_mul_M3} and \ref{alg:mod_mul_risc_v}). Overall, compared to the Montgomery-based implementation, we can utilize the Plantard arithmetic to reduce 128 multiplications in the stack-version pointwise multiplication implementation.
The other two modular reductions, on the other hand, can only be implemented with the Plantard reduction (see Section \ref{subsec:plant_red}). Note that the proposed Plantard reduction is as efficient as the state-of-the-art Montgomery and Barrett reduction on the target platforms.
After the Plantard reduction, the coefficient range of each pointwise multiplication lies in $\plantoutputrange$.

As for the speed-version implementation, we follow the asymmetric multiplication and a better accumulation strategies in \cite{abdulrahman2022faster} and \cite{AbdulrahmanCCHK22}, which help to further minimize the number of modular multiplications or modular reductions, with nearly doubled the stack usage.
The so-called asymmetric multiplication is that during the matrix-vector product $\mathbf{As}$ in Kyber, they \cite{abdulrahman2022faster,AbdulrahmanCCHK22} observed that every row of the NTT-domain matrix $\mathbf{\hat{A}}$ needs to multiply the NTT-domain secret vector $\mathbf{\hat{s}}$ (vector inner product) for $k$ times. For the pointwise multiplication of two NTT-domain polynomials $\hat{a} \circ \hat{s}=\hat{c}$, the term $\hat{s}_{2 i+1} \zeta^{2 \mathrm{br}_7(i)+1}$ also needs to be computed $k$ times throughout the process. 
Therefore, they proposed to reduce these computations by caching $\hat{s}_{2 i+1} \zeta^{2 \mathrm{br}_7(i)+1}$ using an additional polynomial vector $\hat{\mathbf{s}}^{\prime}$. 
When $\mathbf{\hat{s}}$ is multiplied with the first row of matrix $\mathbf{\hat{A}}$, the term $\hat{s}_{2 i+1}\cdot \zeta^{2 \mathrm{br}_7(i)+1} \bmod q$ for each polynomial of the vector $\mathbf{\hat{s}}$ is computed and stored in $\hat{\mathbf{s}}^{\prime}$.   
Similar to the stack-version implementation, this computation can also be speeded up with the efficient Plantard multiplication by a constant. 
The follow-up vector inner products of $\mathbf{\hat{s}}$ with the remaining $k-1$ rows of matrix $\mathbf{\hat{A}}$ can directly load the corresponding $\hat{s}_{2 i+1}\zeta^{2 \mathrm{br}_7(i)+1}$ from $\hat{\mathbf{s}}^{\prime}$ without re-computing the modular multiplication. 

Similarly, the better accumulation technique also trades the speed with extra memory. This technique is utilized in the vector inner product, i.e. $c=\mathbf{\hat{a}}\times \mathbf{\hat{s}}=\sum_{i=0}^{k-1}\hat{a}_i\circ \hat{s}_i,k\in\{2,3,4\}$, which requires to accumulate $k$ pointwise multiplications. Abdulrahman et al. \cite[Section 3.4]{abdulrahman2022faster} showed that one can reduce the final modular reduction of $c_{2i}$ and $c_{2i+1}$, namely only computing $\hat{c}_{2 i}=\hat{a}_{2 i}\cdot \hat{b}_{2 i}+\hat{a}_{2 i+1}\cdot (\hat{b}_{2 i+1} \cdot \zeta^{2 \mathrm{br}_7(i)+1}\bmod q)$ and $\hat{c}_{2 i+1}=\hat{a}_{2 i}\cdot \hat{b}_{2 i+1}+\hat{b}_{2 i}\cdot \hat{a}_{2 i+1}$, and then accumulate these $k$ pointwise multiplications together in a 32-bit array with 256 entries without overflowing 32-bit signed integers. The $k$-th pointwise multiplication will finally reduce each 32-bit coefficient down to 16-bit using the Plantard reduction. After that, the coefficient range of the vector inner product lies in $\plantoutputrange$.
 



\subsubsection{Memory optimizations}\label{subsubsec:memory}
The speed-version Kyber implementation from \cite{abdulrahman2022faster} utilizes the asymmetric multiplication and better accumulation techniques in \cite{AbdulrahmanCCHK22} to minimize the number of modular multiplications or modular reductions.
However, these techniques double the stack usage and pose a challenge for its deployment on memory-constrained IoT devices. In this paper, we propose two memory optimized techniques for the speed-version implementation to address this concern and make it more feasible for such devices. 

The first memory optimized technique is carried out for the asymmetric multiplication technique. 
Recall that Abdulrahman et al. \cite{abdulrahman2022faster,AbdulrahmanCCHK22} proposed to cache the term $\hat{s}_{2 i+1} \zeta^{2 \mathrm{br}_7(i)+1}$ using an additional polynomial vector $\hat{\mathbf{s}}^{\prime}$. 
To make use of the SIMD instruction \textbf{smuad} on Cortex-M4, they also stored a redundant $\hat{s}_{2 i}$ together with each $\hat{s}_{2 i+1} \zeta^{2 \mathrm{br}_7(i)+1}$ in $\hat{\mathbf{s}}^{\prime}$. This leads to the increment of memory usage, as the redundant $\hat{s}_{2 i}$ term has already existed in the original polynomial vector $\mathbf{\hat{s}}$. In this work, we show that the extra memory space is unnecessary on low-end 32-bit platforms. We propose using \textit{poly\_half} and \textit{polyvec\_half} to cache $\hat{s}_{2 i+1} \zeta^{2 \mathrm{br}_7(i)+1}$ only, where each \textit{poly\_half} consists of 128 instead of 256 coefficients and each \textit{polyvec\_half} consists of $k$ \textit{poly\_half}. Our approach results in a halving of stack usage to cache the temporary terms. Specifically, by using the \textit{polyvec\_half} to cache all terms $\hat{s}_{2 i+1} \zeta^{2 \mathrm{br}_7(i)+1}$ for one secret vector in the speed-version code, we could reduce the stack usage by 0.5KiB, 0.75KiB, and 1KiB, in three PKC protocols of Kyber512, Kyber768, and Kyber1024, respectively.

In addition to the memory optimized technique described above, we propose another approach to reduce the memory usage for the better accumulation technique used in the IND-CPA key generation protocol of Kyber, which utilizes a temporary 32-bit array with 256 entries to store the intermediate accumulation result of the vector inner product. 
We observe that in the IND-CPA key generation protocol, the memory space reserved for the 32-bit temporary array and polynomial \textit{pkp} can be merged, where \textit{pkp} is a temporary polynomial in the key generation protocol. 
We introduce a new data structure called \textit{poly\_double} that can hold 512 16-bit coefficients, which has the same memory footprint as the temporary 32-bit array. 
By leveraging data type conversion, we can re-purpose the memory allocated for the temporary array to store \textit{pkp}, reducing its memory usage by 0.5KiB in the key generation protocol.

Together, these two memory optimizations improve the feasibility of the speed-version implementation on memory-constrained IoT devices significantly, such as the selected 16KiB SiFive board. Notably, while the second approach only applies to the KEM key generation protocol, the first technique can be applied to all three KEM protocols of Kyber.

\subsection{Extensibility and Security}
Although this paper primarily focuses on the efficient implementations of Kyber on Cortex-M3 and RISC-V, the Plantard arithmetic with enlarged input range, as described in Algorithm \ref{alg:plant_extend}, is not limited to Kyber. The proposed Plantard arithmetic can be applied to other LBC schemes with 16-bit odd moduli, e.g., NewHope \cite{NewHope} and NTTRU \cite{DBLP:journals/tches/LyubashevskyS19} etc.. 
The optimization techniques proposed in Section \ref{sec:plantard_impl} provide several solutions for the efficient implementation of Plantard arithmetic. These techniques are able to extend to other 32-bit platforms that share similar instruction sets, making it possible to supersede the Montgomery and Barrett arithmetic by the Plantard arithmetic and further speed up LBC.
Besides, most of the optimizations for NTT/INTT can be extended to other LBC schemes with 16-bit NTT, with the exception that the lazy reduction strategy may need to redesign. This is because the Plantard arithmetic may have different input ranges for different moduli used in other LBC schemes.
Last but not least, the memory optimizations proposed for the speed-version Kyber implementation can also be extended to other low-end IoT platforms with similar memory constraints.

However, the proposed optimizations cannot be directly applied to the 32-bit NTT in Dilithium on 32-bit platforms. The Plantard arithmetic for 32-bit modulus would require computing $32\times64$-bit multiplication, which must be separately computed with at least two instructions on 32-bit platforms and thus wipe out the gains of the Plantard arithmetic. However, we believe our optimizations could be applied to Dilithium on 64-bit platforms that can efficiently implement the $32\times 64$-bit multiplication. Moreover, recent research \cite{abdulrahman2022faster} shows that the computation of $c\textbf{s}_i$ in Dilithium can be implemented with 16-bit NTT. Therefore, the proposed optimizations are also applicable to these operations on 32-bit platforms.

In order to resist side-channel attack, our implementation avoids the use of non-constant-time instructions on Cortex-M3. Additionally, we ensure that all the secret-related operations in our implementation are constant-time on both platforms, and our implementation does not leak more sensitive information compared to the reference work, making it secure against simple power analysis and timing side-channel attacks.
\section{Results and Comparisons}\label{sec:results}
This section presents and discusses the experimental results of Kyber on the ARM Cortex-M3 and two RISC-V platforms: the SiFive Freedom E310 board and PQRISCV. All the results on the same platform are obtained under the same environment.

\begin{table*}[tb]
  \caption{Cycle counts of the core polynomial arithmetic, namely NTT, INTT and base multiplication, in Kyber on Cortex-M3, SiFive Freedom E310, and PQRISCV.}
  \label{table:ntt_result}%
  \centering
  \begin{adjustbox}{max width=\textwidth}
    \begin{tabular}{c|c|c|c|c}
    \hline
    {Platform}      & Implementation & NTT &  INTT & Base Multiplication  \\
    \hline
    \multirow{5}[5]{*}{\tabincell{c}{Cortex-M3}}      & Denisa et al. \cite{DBLP:journals/tches/GreconiciKS21} & 10\,874 & 13\,049  & 4\,821 \\ 
       & This work (stack)  & 8\,026  & 8\,594/8\,799   & 4\,311 \\ 
      & This work (speed)  & 8\,026  & 8\,594          & 3\,028/3\,922/5\,851 \\
      & Speedup (stack)   & 26.19\% & 34.14\%/32.57\% & 1.06\% \\
      & Speedup (speed)   & 26.19\% & 34.14\%         & 37.19\%/18.65\%/-21.37\% \\ 
    \hline
    \multirow{5}[5]{*}{\tabincell{c}{SiFive Freedom E310}}  & Denisa et al. \cite{greconici2020kyber}  & 24\,353    & 36\,513    & -$^\text{a}$ \\ 
     & This work (stack)   & 15\,888     & 15\,719/16\,227  & 10\,020 \\
     & This work (speed) &  15\,888   & 15\,719      &  4\,893/5\,662/9\,313  \\ 
      & Speedup (stack)   & 34.76\%   & 56.95\%/55.53\%      & -  \\
    & Speedup (speed)  & 34.76\%   & 56.95\%      & -  \\
    \hline
    \multirow{5}[5]{*}{\tabincell{c}{PQRISCV}}  & Denisa et al. \cite{greconici2020kyber}  &  28\,417   & 42\,636    & -$^\text{a}$ \\ 
    & This work (stack)   & 21\,975     & 23\,666/24\,146  & 12\,236 \\
    & This work (speed) &  21\,975   & 23\,666      &  7\,747/9\,795/13\,068  \\ 
    & Speedup (stack)   & 22.67\%   & 44.49\%/43.37\%      & -  \\
    & Speedup (speed)  & 22.67\%   & 44.49\%      & -  \\
    \hline
    \end{tabular}%
  \end{adjustbox}\\
  \footnotesize{a. \cite{greconici2020kyber} did not provide results for base multiplication.}
\end{table*}%

\subsection{Experimental Setup}\label{sec:setup}
\paragraph*{Cortex-M3 setup} For the Cortex-M3 platform, we utilize the ATSAM3X8E microcontroller on an Arduino Due board \cite{due2017arduino} as the testbed. Our implementation is based on the PQM3 repository \cite{PQM3}, and the GCC v10.2.1 is used to compile the code. We leverage the hardware random number generators available on Cortex-M3 to obtain the necessary random numbers for the Kyber implementation. Stack usage is measured following the methodology outlined in \cite{PQM3}. Additionally, SHA3 and SHAKE are implemented using the optimized assembly Keccak permutation token from the eXtended Keccak Code Package (XKCP)\footnote{\url{https://github.com/XKCP/XKCP}}. 

\paragraph*{RISC-V setup} The realistic RISC-V platform we choose is the SiFive Freedom E310 board, which is equipped with a 32-bit E31 RISC-V core \cite{SiFive:Manual}. Due to its limited 16KiB memory, a direct deployment of Kyber on this platform is infeasible. Hence, in order to provide detailed comparisons with the existing Kyber implementation on RISC-V, we also present experimental results obtained from the RISC-V simulator used in PQRISCV \cite{PQRISCV}, which uses a RISC-V CPU implemented in pqriscv-vexriscv\footnote{https://github.com/mupq/pqriscv-vexriscv} as the PQC benchmarking platform. 
The GCC compiler we used for RISC-V is the RISC-V GNU v10.2.0. Since both of the RISC-V platforms do not have hardware random number generators, we reuse the function in PQRISCV \cite{PQRISCV} to generate the random numbers in Kyber. 
Furthermore, for the implementation of SHA3 and SHAKE, we rely on the optimized Keccak permutation provided by Ko \cite{stoffelen2019efficient} in PQRISCV. 

To support the Kyber-90s variants, we adopt the optimized AES implementation presented in \cite{adomnicai2020fixslicing} on Cortex-M3 and PQRISCV similar to the PQM3 \cite{PQM3}. 
On the SiFive platform, however, we use the reference SHA3, SHAKE, and AES implementations provided in SUPERCOP\footnote{\url{http://bench.cr.yp.to/supercop.html}} to reduce the code size. This is because a bloated code size might impact the performance of the platform. Further discussion on this issue will be provided in Section \ref{sec:res_scheme}.


\subsection{Performance of the Polynomial Arithmetic}\label{sec:res_arith}

Table \ref{table:ntt_result} presents the performance comparison of the polynomial arithmetic. The state-of-the-art Kyber implementation on Cortex-M3 was presented by Denisa et al. \cite{DBLP:journals/tches/GreconiciKS21}, which is also the latest implementation integrated in PQM3 \cite{PQM3}. 
On RISC-V, the closest related work also comes from Denisa et al. \cite{greconici2020kyber}. 
However, since the implementations on RISC-V \cite{greconici2020kyber} do not involve any memory optimization techniques, we are unable to deploy their Kyber implementation on the SiFive platform. Instead, we only incorporate their NTT and INTT implementations on the SiFive platform for comparison.

As mentioned in Section \ref{subsubsec:lazy_reduction}, we eliminate the modular reduction of coefficients for NTT, therefore, both the stack-version and speed-version NTT cost the same number of cycles. Specifically, the proposed NTT implementation obtains speedups of 26.19\%, 34.76\%, and 22.67\% on the ARM Cortex-M3, SiFive Freedom E310, and PQRISCV, respectively, compared to the implementations in \cite{DBLP:journals/tches/GreconiciKS21} and \cite{greconici2020kyber}. In terms of INTT, because we manage to eliminate the modular reduction of coefficients in the stack-version INTT of Kyber512 and the speed-version INTT of all three Kyber variants, the INTT of Kyber512 is slightly faster than the one in the stack-version INTT of Kybe768 and Kyber1024. Overall, our INTT achieve speedups of 32.57\%/34.14\%,55.53\%/ 56.95\%, and 43.36\%/44.49\% on the ARM Cortex-M3, SiFive Freedom E310, and PQRISCV, respectively, compared to the implementations in \cite{DBLP:journals/tches/GreconiciKS21} and \cite{greconici2020kyber}. 

Huang et al. \cite{huang2022improved} reported that the stack-version base multiplication with the Plantard arithmetic is slower than the previous Montgomery-based counterpart on Cortex-M4 due to the lack of register usage resulting from the Plantard multiplication of two variables. However, on Cortex-M3 and RISC-V platforms, the stack-version base multiplication does not have this issue thus could even benefit from the Plantard multiplication by the twiddle factor. 
The speed-version implementation offers three variants of base multiplication to deploy the asymmetric multiplication and better accumulation techniques described in Section \ref{subsubsec:basemul}, which help to mitigate the need of some modular reductions.
Therefore, two of the three variants are 29.74\% and 9.02\% faster than our stack-version base multiplication, while the third variant is 35.72\% slower, mainly due to additional loading cycles and high register pressure.
Compared to the state-of-the-art Montgomery-based implementation in in \cite{DBLP:journals/tches/GreconiciKS21}, these two variants of the speed-version base multiplications are 37.19\% and 18.65\% faster than the base multiplication due to the use of the improved Plantard arithmetic, asymmetric multiplication and better accumulation techniques. Since the implementation from Denisa et al. \cite{greconici2020kyber} did not provide results for the base multiplication, we do not compare it with our results on the RISC-V platforms.
Overall, the speed-version implementation offers faster INTT and base multiplications for all variants of Kyber than the stack-version. 
It should be noted that the slower variant of base multiplication will only be used at the final stage of the innver-vector product  and matrix-vector product while the two faster variants will be used repeatedly in these two operations. Therefore, the faster INTT and base multiplications would speed up these two time-consuming operations. Together with the proposed memory optimizations in Section \ref{subsubsec:memory}, the speed-version implementation provides an effective time-memory trade-off for Kyber on low-end 32-bit platforms. Detailed improvements of the speed-version implementation could be found in Table \ref{table:kem_result}. 


\begin{table*}[tb]
  \caption{Cycle counts (cc) and stack usage (Bytes) of KeyGen, Encaps, and Decaps on Cortex-M3, SiFive Freedom E310, and PQRISCV. The first row of each entry indicates the cycle count and the second row refers to stack usage.}
  \label{table:kem_result}%
  \centering
  \begin{adjustbox}{max width=\textwidth}
    \begin{tabular}{c|c|c|c|c|c|c|c|c|c|c}
    \hline 
    \multirow{2}{*}{Platform}      & \multirow{2}{*}{Implementation} & \multicolumn{3}{c|}{Kyber512} &  \multicolumn{3}{c|}{Kyber768} & \multicolumn{3}{c}{Kyber1024} \\
    \cline{3-11} &                     & KeyGen  & Encaps  & Decaps   & KeyGen  & Encaps  & Decaps  & KeyGen  & Encaps  & Decaps  \\
    \hline
    \multirow{6}{*}{\tabincell{c}{Cortex-M3}} & 
     \multirow{2}{*}{\tabincell{c}{Denisa et al.\cite{DBLP:journals/tches/GreconiciKS21}}}           & 541k & 650k  & 622k & 878k    & 1\,054k   & 1\,010k & 1\,388k  & 1\,602k  & 1\,543k \\
    & & 2\,212 & 2\,300 & 2\,308 & 3\,084 & 2\,772 & 2\,788 &  3\,596 & 3\,284 & 3\,300\\
    \cline{2-11}    
    & \multirow{2}{*}{\tabincell{c}{This work (stack)}}  &  519k  & 628k  & 590k &  844k & 1\,025k  & 967k &  1\,342k  & 1\,563k  & 1\,486k \\
    & & 2\,212 & 2\,300 & 2\,308 & 3\,084 & 2\,772 & 2\,788 &  3\,596 & 3\,284 & 3\,300\\
    \cline{2-11}
    & \multirow{2}{*}{\tabincell{c}{This work (speed)}}      & 518k & 626k  & 587k & 842k  & 1\,017k  & 958k & 1\,333k  & 1\,548k  & 1\,471k \\
    & &  3\,268 & 3\,860 & 3\,860 & 4\,044  & 4\,636  & 4\,636 & 4\,812  & 5\,404  & 5\,404 \\
    \hline
    \multirow{6}{*}{\tabincell{c}{PQRISCV}}  
    & \multirow{2}{*}{\tabincell{c}{Denisa et al.\cite{greconici2020kyber}}}           & 2\,229k   & 2\,927k   &  2\,856k & 4\,166k   & 5\,071k   &  4\,957k & 6\,696k   & 7\,809k  & 7\,662k\\
    & &  6\,544  & 9\,200  & 9\,984 & 10\,640  &  13\,808 & 14\,944 &   15\,760 &  19\,440 & 21\,056 \\
    \cline{2-11}  
    & \multirow{2}{*}{\tabincell{c}{This work (stack)}}  &  1\,937k   &  2\,355k & 2\,100k & 3\,147k  & 3\,822k  & 3\,467k  & 4\,964k   & 5\,794k  & 5\,344k \\
    & & 2\,408  & 2\,488  & 2\,520 & 2\,952  & 3\,016  & 3\,032 &  3\,464  & 3\,528  & 3\,544 \\
    \cline{2-11}
    & \multirow{2}{*}{\tabincell{c}{This work (speed)}}     & 1\,926k   & 2\,339k  & 2\,084k &  3\,104k & 3\,768k  & 3\,413k  & 4\,890k   & 5\,704k  & 5\,254k\\
    & &  3\,432  & 4\,024  & 4\,040 & 4\,216  & 4\,808  & 4\,840 &  5\,032  & 5\,608  & 5\,656\\
    \hline
    \multirow{4}{*}{\tabincell{c}{SiFive Freedom E310}} & 
    \multirow{2}{*}{\tabincell{c}{This work (stack)}}  &  1\,497k   &  1\,812k & 1\,601k & 2\,413k  & 2\,929k  & 2\,635k  & 3\,794k   & 4\,435k  & 4\,045k \\
    & & 2\,580   & 2\,660  & 2\,708 & 3\,060  & 3\,124  & 3\,156 &  3\,572  & 3\,636  & 3\,668 \\
    \cline{2-11}
    & \multirow{2}{*}{\tabincell{c}{This work (speed)}}     & 1\,597k   & 1\,903k  & 1\,674k &  2\,731k & 3\,203k  & 2\,919k  & -   & -  & -\\
    & &  3\,620  & 4\,212  & 4\,244 & 4\,340  & 4\,932  & 4\,964 &  -  & -  & - \\
    \hline\hline 
    \multirow{2}{*}{Platform}      & \multirow{2}{*}{Implementation} & \multicolumn{3}{c|}{Kyber512-90s} &  \multicolumn{3}{c|}{Kyber768-90s} & \multicolumn{3}{c}{Kyber1024-90s} \\
    \cline{3-11} &                     & KeyGen  & Encaps  & Decaps   & KeyGen  & Encaps  & Decaps  & KeyGen  & Encaps  & Decaps  \\
    \hline
    \multirow{6}{*}{\tabincell{c}{Cortex-M3}} & 
     \multirow{2}{*}{\tabincell{c}{Denisa et al.\cite{DBLP:journals/tches/GreconiciKS21}}}           & 467k & 527k  & 558k & 782k    & 868k   & 909k & 1\,224k  & 1\,332k  & 1\,383k \\
    & & 2\,904 & 2\,992 & 3\,000 & 3\,432 & 3\,504 & 3\,512 &  4\,636 & 4\,000 & 4\,016\\
    \cline{2-11}    
    & \multirow{2}{*}{\tabincell{c}{This work (stack)}}  &  445k  & 505k  & 526k &  748k & 839k  & 865k &  1\,178k  & 1\,293k  & 1\,326k \\
    & & 2\,904 & 2\,992 & 3\,000 & 3\,432 & 3\,504 & 3\,512 &  4\,636 & 4\,000 & 4\,016\\
    \cline{2-11}
    & \multirow{2}{*}{\tabincell{c}{This work (speed)}}      & 443k & 502k  & 523k & 741k  & 828k  & 854k & 1\,170k  & 1\,274k  & 1\,306k \\
    & &  4\,000 & 4\,592 & 4\,592 & 4\,776  & 5\,368  & 5\,368 & 5\,544  & 6\,136  & 6\,136 \\
    \hline
    \multirow{6}{*}{\tabincell{c}{PQRISCV}}  
    & \multirow{2}{*}{\tabincell{c}{Denisa et al.\cite{greconici2020kyber}}}           & 3\,042k   & 3\,491k   &  3\,627k & 6\,106k   & 6\,677k   &  6\,854k & 10\,246k   & 10\,953k  & 11\,181k\\
    & &  6\,656  & 9\,312  & 10\,096 & 10\,752  &  13\,920 & 15\,056 &   15\,872 &  19\,552 & 21\,168 \\
    \cline{2-11}  
    & \multirow{2}{*}{\tabincell{c}{This work (stack)}}  & 2\,667k   & 2\,902k & 2\,854k &  5,126k  &  5,434k  &  5,370k  & 8\,578k   &  8\,983k  &  8\,908k \\
    & & 2\,704  & 2\,616  & 2\,648 & 3\,248  & 3\,144  & 3\,160 &  3\,760  & 3\,656  & 3\,672 \\
    \cline{2-11}
    & \multirow{2}{*}{\tabincell{c}{This work (speed)}}     & 2\,654k   & 2\,836k  & 2\,788k &   5\,081k & 5\,379k  &  5\,314k  &  8,499k   &  8,888k  &  8,813k\\
    & &  3\,744  & 4\,216  & 4\,232 & 4\,528  & 5\,000  & 5\,032 &  5\,296   & 5\,752   & 5\,800\\
    \hline
    \multirow{4}{*}{\tabincell{c}{SiFive Freedom E310}} & 
    \multirow{2}{*}{\tabincell{c}{This work (stack)}}  &  8\,326k   &  8\,619k & 8\,884k & 11\,443k  & 11\,758k  & 12\,061k  &  -  & -  & - \\
    & & 2\,780   & 2\,860  & 2\,908 & 3\,292  & 3\,356  & 3\,388 &  -  & -  & - \\
    \cline{2-11}
    & \multirow{2}{*}{\tabincell{c}{This work (speed)}}     & 8\,438k  & 8\,757k  & 9\,046k &  - & -  & -  & -   & -  & -\\
    & &  3\,820  & 4\,412  & 4\,444 & -  & - & - &  -  & -  & - \\
    \hline
    \end{tabular}
  \end{adjustbox}
  \end{table*}%

\subsection{Performance and Stack Usage of Kyber}\label{sec:res_scheme}
Table \ref{table:kem_result} presents the cycle counts of the KEM protocols for Kyber and Kyber-90s, including key generation (KeyGen), encapsulation (Encaps), and decapsulation (Decaps), which are obtained by repeating each protocol one hundred times and then computing the average results. The stack usage is measured in a similar way to \cite{PQM3} and \cite{greconici2020kyber}. We have implemented the stack-version and speed-version of Kyber on Cortex-M3 and RISC-V platforms similar to the optimized implementation on Cortex-M4 \cite{abdulrahman2022faster}. Our stack-version implementation on Cortex-M3 offers speedups of 3.38\%$\sim$5.14\%, 2.75\%$\sim$4.26\%, and 2.43\%$\sim$3.69\% compared to \cite{DBLP:journals/tches/GreconiciKS21} for Kyber512, Kyber768, and Kyber1024, respectively, while having the same stack usage as theirs. Similarly, our speed-version implementation on Cortex-M3 outperforms the work in \cite{DBLP:journals/tches/GreconiciKS21} with a speedup of 3.69\%$\sim$5.63\%, 3.51\%$\sim$5.15\%, and 3.37\%$\sim$4.67\% for Kyber512, Kyber768, and Kyber1024, respectively, with 1.31$\sim$1.68 times stack usage compared to their implementation. Notably, thanks to the proposed memory optimizations in \ref{subsubsec:memory}, our speed-version Kyber implementation on Cortex-M3 reduces the stack usage by 23.50\%$\sim$28.31\% compared to its counterpart on Cortex-M4 \cite[Table 4]{abdulrahman2022faster}.

To the best of our knowledge, there exists only one Kyber implementation on RISC-V (cf. Denisa et al. \cite{greconici2020kyber}). They have provided optimized assembly implementation for NTT/INTT of Kyber and leave rooms for further exploration to optimize memory footprints using memory optimized techniques.
Table \ref{table:kem_result} shows that their implementation has a stack usage that is 2.71$\sim$5.94 times larger than our stack-version implementation on PQRISCV. Notably, our stack-version implementation achieves a speedup of 13.10\%$\sim$26.47\%, 24.46\%$\sim$30.05\%, and 25.80\%$\sim$30.25\% for the three variants of Kyber, respectively, compared to their implementation. Moreover, our speed-version implementation outperforms theirs by 13.59\%$\sim$27.03\%, 25.49\%$\sim$31.15\%, and 26.96\%$\sim$31.43\% using only 26.86\%$\sim$52.44\% of their stack usage for the three variants of Kyber, respectively. It can be deduced that the large speedups of our implementations compared to \cite{greconici2020kyber} mainly stems from the fact that their implementation is not as optimized and does not incorporate the recent optimization strategies described in the recent literature \cite{botros2019memory,alkim2020cortex,abdulrahman2022faster}.

The extremely large stack usage of Denisa et al.'s implementation \cite{greconici2020kyber} makes it infeasible to deploy on the 16KiB SiFive board. Hence, we only provide results for the proposed optimized implementations on this platform. Our results demonstrate that with the memory optimization strategies proposed in Section \ref{subsubsec:memory}, it is possible to deploy the speed-version of Kyber512 and Kyber768 on the selected SiFive board. Notably, these two variants of Kyber are unable to be deployed on this platform before using our memory optimizations. However, since the speed-version implementation has a larger code size than the stack-version, for example, 13.75KiB versus 12.44KiB for Kyber512, it is slower than the stack-version on the SiFive board. 
This can be attributed to the fact that the larger code size of the speed-version implementation spills more instructions into the ROM, leading to much slower performance compared to the stack-version, as stated in  \cite[Section 5.3]{jipeng2021}. However, as demonstrated in our Cortex-M3 and PQRISCV results, the speed-version implementation can still outperform the stack-version on different platforms with different characteristics. Although the performance of the speed-version implementation on the SiFive board is slower than the stack-version, we show the feasibility of deploying both stack-version and speed-version Kyber implementations on memory-constrained IoT devices, except for the speed-version of Kyber1024. This is made possible by the memory optimized strategies proposed in Section \ref{subsubsec:memory}.

As for the Kyber-90s variants, we yeild $2.93\%\sim 6.27\%$ speed-ups on Cortex-M3 and $12.33\%\sim 23.13\%$ speed-ups on PQRISCV. The speed improvements are very similar to the Kyber variants on Cortex-M3 and PQRISCV. 
On the SiFive platform, the adoption of the reference AES implementation results in the Kyber-90s variants being over $3\times$ slower than the Kyber variants. 
We did not utilize the optimized AES implementation in \cite{adomnicai2020fixslicing} because it would increase the code size, leading to slower performance. 
Additionally, the reference AES implementation introduces additional stack usage, posing challenges on the deployment of the speed-version of Kyber768-90s and both versions of Kyber1024-90s on the selected memory-constrained SiFive platforms. Therefore, we are unable to provide their results on the SiFive platform. Addressing these deployment issues for Kyber-90s will require additional efforts and optimizations. However, we believe that if the target platforms have AES hardware support, the Kyber-90s variants will remain a promising option on these platforms.
\section{Conclusions}\label{sec:conclusions}
This paper presents faster implementations of Kyber on low-end 32-bit IoT devices, specifically for ARM Cortex-M3 and RISC-V. In particular, we prove that the input range of the Plantard arithmetic can be further enlarged. The enlarged Plantard arithmetic could also be tailored for Kyber's modulus. Leveraging this theoretical foundation, we present efficient implementations of the Plantard arithmetic on the aforementioned devices by exploiting their specific ISA characteristics. We then propose various optimized strategies to improve the efficiency of NTT/INTT. Furthermore, two memory optimized techniques are introduced for the speed-version Kyber implementation, making it more feasible on low-end IoT devices. Overall, we achieve new speed records for Kyber on these low-end 32-bit platforms.

\section*{Acknowledgments}
This work is partially supported by the 
National Key Research and Development Program of China (2022YFB2702000), 
National Natural Science Foundation of China (62002023, 62002239, 62372417, and 62071222), Guangdong Provincial Key Laboratory IRADS (2022B1212010006, R0400001-22), Guangdong Province General Universities Key Field Project (New Generation Information Technology) (2023ZDZX1033), Zhejiang Lab Open Research Project (K2022PD0AB03),
Jiangsu Province 100 Foreign Experts Introduction Plan (BX2022012), T{\"U}B{\.I}TAK Projects (2232-118C332 and 1001-121F348), 
ITF Project ITS/098/22, and the InnoHK Project CIMDA.

\bibliographystyle{IEEEtran}
\bibliography{IEEEabrv,crypto}

\begin{thebibliography}{10}
\providecommand{\url}[1]{#1}
\csname url@samestyle\endcsname
\providecommand{\newblock}{\relax}
\providecommand{\bibinfo}[2]{#2}
\providecommand{\BIBentrySTDinterwordspacing}{\spaceskip=0pt\relax}
\providecommand{\BIBentryALTinterwordstretchfactor}{4}
\providecommand{\BIBentryALTinterwordspacing}{\spaceskip=\fontdimen2\font plus
\BIBentryALTinterwordstretchfactor\fontdimen3\font minus
  \fontdimen4\font\relax}
\providecommand{\BIBforeignlanguage}[2]{{%
\expandafter\ifx\csname l@#1\endcsname\relax
\typeout{** WARNING: IEEEtran.bst: No hyphenation pattern has been}%
\typeout{** loaded for the language `#1'. Using the pattern for}%
\typeout{** the default language instead.}%
\else
\language=\csname l@#1\endcsname
\fi
#2}}
\providecommand{\BIBdecl}{\relax}
\BIBdecl

\bibitem{gidney2021factor}
\BIBentryALTinterwordspacing
C.~Gidney and M.~Eker{\aa}, ``How to factor 2048 bit {RSA} integers in 8 hours
  using 20 million noisy qubits,'' \emph{Quantum}, vol.~5, p. 433, 2021.
  [Online]. Available: \url{https://doi.org/10.22331/q-2021-04-15-433}
\BIBentrySTDinterwordspacing

\bibitem{NIST:PQCStandard}
NIST, ``Announcing {PQC} candidates to be standardized, plus fourth round
  candidates,''
  \url{https://csrc.nist.gov/News/2022/pqc-candidates-to-be-standardized-and-round-4},
  Jul. 2022, (accessed Mar. 13, 2023).

\bibitem{DBLP:conf/eurosp/BosDKLLSSSS18}
\BIBentryALTinterwordspacing
J.~W. Bos, L.~Ducas, E.~Kiltz, T.~Lepoint, V.~Lyubashevsky, J.~M. Schanck,
  P.~Schwabe, G.~Seiler, and D.~Stehl{\'{e}}, ``{CRYSTALS} - {Kyber}: {A}
  {CCA-Secure} module-lattice-based {KEM},'' in \emph{2018 {IEEE} European
  Symposium on Security and Privacy, EuroS{\&}P 2018, London, United Kingdom,
  April 24-26, 2018}.\hskip 1em plus 0.5em minus 0.4em\relax {IEEE}, 2018, pp.
  353--367. [Online]. Available:
  \url{https://doi.org/10.1109/EuroSP.2018.00032}
\BIBentrySTDinterwordspacing

\bibitem{Dilithium}
\BIBentryALTinterwordspacing
L.~Ducas, E.~Kiltz, T.~Lepoint, V.~Lyubashevsky, P.~Schwabe, G.~Seiler, and
  D.~Stehl{\'{e}}, ``Crystals-dilithium: {A} lattice-based digital signature
  scheme,'' \emph{{IACR} Trans. Cryptogr. Hardw. Embed. Syst.}, vol. 2018,
  no.~1, pp. 238--268, 2018. [Online]. Available:
  \url{https://doi.org/10.13154/tches.v2018.i1.238-268}
\BIBentrySTDinterwordspacing

\bibitem{prest2020falcon}
T.~Prest, P.-A. Fouque, J.~Hoffstein, P.~Kirchner, V.~Lyubashevsky, T.~Pornin,
  T.~Ricosset, G.~Seiler, W.~Whyte, and Z.~Zhang, ``Falcon,''
  \emph{Post-Quantum Cryptography Project of NIST}, 2020.

\bibitem{sphincs+}
\BIBentryALTinterwordspacing
D.~J. Bernstein, A.~H{\"{u}}lsing, S.~K{\"{o}}lbl, R.~Niederhagen,
  J.~Rijneveld, and P.~Schwabe, ``The {SPHINCS}\({}^{\mbox{+}}\) signature
  framework,'' in \emph{Proceedings of the 2019 {ACM} {SIGSAC} Conference on
  Computer and Communications Security, {CCS} 2019, London, UK, November 11-15,
  2019}, L.~Cavallaro, J.~Kinder, X.~Wang, and J.~Katz, Eds.\hskip 1em plus
  0.5em minus 0.4em\relax {ACM}, 2019, pp. 2129--2146. [Online]. Available:
  \url{https://doi.org/10.1145/3319535.3363229}
\BIBentrySTDinterwordspacing

\bibitem{Statista:IoT_predict}
Statista, ``{IoT} connected devices worldwide 2019-2030,''
  \url{https://www.statista.com/statistics/1183457/iot-connected-devices-worldwide/},
  Jul. 2023, (accessed Mar. 13, 2023).

\bibitem{PQM4}
\BIBentryALTinterwordspacing
M.~J. Kannwischer, J.~Rijneveld, P.~Schwabe, and K.~Stoffelen, ``{PQM4}:
  {Post-quantum} crypto library for the {ARM} {Cortex-M4},'' {Accessed}: Mar.
  13, 2023. [Online]. Available: \url{https://github.com/mupq/pqm4}
\BIBentrySTDinterwordspacing

\bibitem{NewHope}
\BIBentryALTinterwordspacing
E.~Alkim, L.~Ducas, T.~P{\"{o}}ppelmann, and P.~Schwabe, ``Post-quantum key
  exchange - {A} new hope,'' in \emph{25th {USENIX} Security Symposium,
  {USENIX} Security 16, Austin, TX, USA, August 10-12, 2016}, T.~Holz and
  S.~Savage, Eds.\hskip 1em plus 0.5em minus 0.4em\relax {USENIX} Association,
  2016, pp. 327--343. [Online]. Available:
  \url{https://www.usenix.org/conference/usenixsecurity16/technical-sessions/presentation/alkim}
\BIBentrySTDinterwordspacing

\bibitem{seiler2018faster}
\BIBentryALTinterwordspacing
G.~Seiler, ``Faster {AVX2} optimized {NTT} multiplication for {Ring-LWE}
  lattice cryptography,'' \emph{{IACR} Cryptol. ePrint Arch.}, p.~39, 2018.
  [Online]. Available: \url{http://eprint.iacr.org/2018/039}
\BIBentrySTDinterwordspacing

\bibitem{botros2019memory}
\BIBentryALTinterwordspacing
L.~Botros, M.~J. Kannwischer, and P.~Schwabe, ``Memory-efficient high-speed
  implementation of {Kyber} on {Cortex-M4},'' in \emph{Progress in Cryptology -
  {AFRICACRYPT} 2019 - 11th International Conference on Cryptology in Africa,
  Rabat, Morocco, July 9-11, 2019, Proceedings}, ser. Lecture Notes in Computer
  Science, J.~Buchmann, A.~Nitaj, and T.~Rachidi, Eds., vol. 11627.\hskip 1em
  plus 0.5em minus 0.4em\relax Springer, 2019, pp. 209--228. [Online].
  Available: \url{https://doi.org/10.1007/978-3-030-23696-0\_11}
\BIBentrySTDinterwordspacing

\bibitem{alkim2020cortex}
\BIBentryALTinterwordspacing
E.~Alkim, Y.~A. Bilgin, M.~Cenk, and F.~G{\'{e}}rard, ``Cortex-{M4}
  optimizations for \{R, M\} {LWE} schemes,'' \emph{{IACR} Trans. Cryptogr.
  Hardw. Embed. Syst.}, vol. 2020, no.~3, pp. 336--357, 2020. [Online].
  Available: \url{https://doi.org/10.13154/tches.v2020.i3.336-357}
\BIBentrySTDinterwordspacing

\bibitem{DBLP:journals/tches/GreconiciKS21}
\BIBentryALTinterwordspacing
D.~O.~C. Greconici, M.~J. Kannwischer, and A.~Sprenkels, ``Compact {Dilithium}
  implementations on {Cortex-M3 and Cortex-M4},'' \emph{{IACR} Trans. Cryptogr.
  Hardw. Embed. Syst.}, vol. 2021, no.~1, pp. 1--24, 2021. [Online]. Available:
  \url{https://doi.org/10.46586/tches.v2021.i1.1-24}
\BIBentrySTDinterwordspacing

\bibitem{abdulrahman2022faster}
A.~Abdulrahman, V.~Hwang, M.~J. Kannwischer, and D.~Sprenkels, ``Faster {Kyber}
  and {Dilithium} on the {Cortex-M4},'' in \emph{Applied Cryptography and
  Network Security - 20th International Conference, {ACNS} 2022, Rome, Italy,
  June 20-23, 2022, Proceedings}, ser. Lecture Notes in Computer Science,
  G.~Ateniese and D.~Venturi, Eds., vol. 13269.\hskip 1em plus 0.5em minus
  0.4em\relax Springer, 2022, pp. 853--871.

\bibitem{cooley1965algorithm}
J.~W. Cooley and J.~W. Tukey, ``An algorithm for the machine calculation of
  complex {Fourier} series,'' \emph{Mathematics of computation}, vol.~19,
  no.~90, pp. 297--301, 1965.

\bibitem{gentleman1966fast}
\BIBentryALTinterwordspacing
W.~M. Gentleman and G.~Sande, ``{Fast Fourier Transforms}: for fun and
  profit,'' in \emph{American Federation of Information Processing Societies:
  Proceedings of the {AFIPS} '66 Fall Joint Computer Conference, November 7-10,
  1966, San Francisco, California, {USA}}, ser. {AFIPS} Conference Proceedings,
  vol.~29.\hskip 1em plus 0.5em minus 0.4em\relax {AFIPS} / {ACM} / Spartan
  Books, Washington {D.C.}, 1966, pp. 563--578. [Online]. Available:
  \url{https://doi.org/10.1145/1464291.1464352}
\BIBentrySTDinterwordspacing

\bibitem{montgomery1985modular}
P.~L. Montgomery, ``Modular multiplication without trial division,''
  \emph{Mathematics of computation}, vol.~44, no. 170, pp. 519--521, 1985.

\bibitem{barrett1986implementing}
\BIBentryALTinterwordspacing
P.~Barrett, ``Implementing the {Rivest} {Shamir} and {Adleman} public key
  encryption algorithm on a standard digital signal processor,'' in
  \emph{Advances in Cryptology - {CRYPTO} '86, Santa Barbara, California, USA,
  1986, Proceedings}, ser. Lecture Notes in Computer Science, A.~M. Odlyzko,
  Ed., vol. 263.\hskip 1em plus 0.5em minus 0.4em\relax Springer, 1986, pp.
  311--323. [Online]. Available:
  \url{https://doi.org/10.1007/3-540-47721-7\_24}
\BIBentrySTDinterwordspacing

\bibitem{Plantard:2021:Mod}
\BIBentryALTinterwordspacing
T.~Plantard, ``Efficient word size modular arithmetic,'' \emph{{IEEE}
  Transactions on Emerging Topics in Computing}, vol.~9, no.~3, pp. 1506--1518,
  2021. [Online]. Available: \url{https://doi.org/10.1109/TETC.2021.3073475}
\BIBentrySTDinterwordspacing

\bibitem{huang2022improved}
\BIBentryALTinterwordspacing
J.~Huang, J.~Zhang, H.~Zhao, Z.~Liu, R.~C.~C. Cheung, {\c{C}}.~K. Ko{\c{c}},
  and D.~Chen, ``Improved {Plantard} arithmetic for lattice-based
  cryptography,'' \emph{{IACR} Trans. Cryptogr. Hardw. Embed. Syst.}, vol.
  2022, no.~4, pp. 614--636, 2022. [Online]. Available:
  \url{https://doi.org/10.46586/tches.v2022.i4.614-636}
\BIBentrySTDinterwordspacing

\bibitem{DBLP:journals/dcc/LangloisS15}
\BIBentryALTinterwordspacing
A.~Langlois and D.~Stehl{\'{e}}, ``Worst-case to average-case reductions for
  module lattices,'' \emph{Des. Codes Cryptogr.}, vol.~75, no.~3, pp. 565--599,
  2015. [Online]. Available: \url{https://doi.org/10.1007/s10623-014-9938-4}
\BIBentrySTDinterwordspacing

\bibitem{DBLP:conf/stoc/Regev05}
\BIBentryALTinterwordspacing
O.~Regev, ``On lattices, learning with errors, random linear codes, and
  cryptography,'' in \emph{Proceedings of the 37th Annual {ACM} Symposium on
  Theory of Computing, Baltimore, MD, USA, May 22-24, 2005}, H.~N. Gabow and
  R.~Fagin, Eds.\hskip 1em plus 0.5em minus 0.4em\relax {ACM}, 2005, pp.
  84--93. [Online]. Available: \url{https://doi.org/10.1145/1060590.1060603}
\BIBentrySTDinterwordspacing

\bibitem{DBLP:conf/eurocrypt/LyubashevskyPR10}
\BIBentryALTinterwordspacing
V.~Lyubashevsky, C.~Peikert, and O.~Regev, ``On ideal lattices and learning
  with errors over rings,'' in \emph{Advances in Cryptology - {EUROCRYPT} 2010,
  29th Annual International Conference on the Theory and Applications of
  Cryptographic Techniques, Monaco / French Riviera, May 30 - June 3, 2010.
  Proceedings}, ser. Lecture Notes in Computer Science, H.~Gilbert, Ed., vol.
  6110.\hskip 1em plus 0.5em minus 0.4em\relax Springer, 2010, pp. 1--23.
  [Online]. Available: \url{https://doi.org/10.1007/978-3-642-13190-5\_1}
\BIBentrySTDinterwordspacing

\bibitem{DBLP:conf/crypto/FujisakiO99}
\BIBentryALTinterwordspacing
E.~Fujisaki and T.~Okamoto, ``Secure integration of asymmetric and symmetric
  encryption schemes,'' in \emph{Advances in Cryptology - {CRYPTO} '99, 19th
  Annual International Cryptology Conference, Santa Barbara, California, USA,
  August 15-19, 1999, Proceedings}, ser. Lecture Notes in Computer Science,
  M.~J. Wiener, Ed., vol. 1666.\hskip 1em plus 0.5em minus 0.4em\relax
  Springer, 1999, pp. 537--554. [Online]. Available:
  \url{https://doi.org/10.1007/3-540-48405-1\_34}
\BIBentrySTDinterwordspacing

\bibitem{avanzi2020crystals}
R.~Avanzi, J.~Bos, L.~Ducas, E.~Kiltz, T.~Lepoint, V.~Lyubashevsky, J.~M.
  Schanck, P.~Schwabe, G.~Seiler, and D.~Stehl{\'e}, ``{CRYSTALS}-{Kyber}
  algorithm specifications and supporting documentation,'' 2020.

\bibitem{AbdulrahmanCCHK22}
\BIBentryALTinterwordspacing
A.~Abdulrahman, J.~Chen, Y.~Chen, V.~Hwang, M.~J. Kannwischer, and B.~Yang,
  ``Multi-moduli {NTTs} for {Saber} on {Cortex-M3 and Cortex-M4},''
  \emph{{IACR} Trans. Cryptogr. Hardw. Embed. Syst.}, vol. 2022, no.~1, pp.
  127--151, 2022. [Online]. Available:
  \url{https://doi.org/10.46586/tches.v2022.i1.127-151}
\BIBentrySTDinterwordspacing

\bibitem{due2017arduino}
\BIBentryALTinterwordspacing
A.~Due and A.~Core, ``Arduino due,'' \emph{Retrieved}, Jul. 2016. [Online].
  Available: \url{https://www.arduino.cc/en/Main/ArduinoBoardDue}
\BIBentrySTDinterwordspacing

\bibitem{de2015performance}
W.~de~Groot, ``A performance study of {X25519} on {Cortex-M3} and {M4},'' Ph.D.
  dissertation, Eindhoven University of Technology Eindhoven, The Netherlands,
  2015.

\bibitem{SiFive:Manual}
SiFive, ``Sifive {FE310-G002} {Manual},''
  \url{https://sifive.cdn.prismic.io/sifive/b56b304f-cd2d-421b-9c14-6b35c33f172e_fe310-g002-manual-v1p4.pdf},
  (accessed Mar. 13, 2023).

\bibitem{yang2024modular}
Y.~Yang, Y.~Jia, and G.~Xu, ``On modular algorithms and butterfly operations in
  number theoretic transform,'' \emph{arXiv preprint arXiv:2402.00675}, 2024.

\bibitem{cryptoeprint:2022/956}
\BIBentryALTinterwordspacing
J.~Huang, J.~Zhang, H.~Zhao, Z.~Liu, R.~C.~C. Cheung, {\c{C}}.~K. Ko{\c{c}},
  and D.~Chen, ``Improved plantard arithmetic for lattice-based cryptography,''
  Cryptology ePrint Archive, Paper 2022/956, 2022,
  \url{https://eprint.iacr.org/2022/956}. [Online]. Available:
  \url{https://eprint.iacr.org/2022/956}
\BIBentrySTDinterwordspacing

\bibitem{greconici2020kyber}
D.~Greconici, ``Kyber on {RISC-V},'' Master's thesis, Radboud University
  Nijmegen, The Netherlands, 2020.

\bibitem{PQM3}
\BIBentryALTinterwordspacing
PQM3, ``{PQM3}: {Post-quantum} crypto library for the {ARM} {Cortex-M3},''
  {Accessed}: Mar. 13, 2023. [Online]. Available:
  \url{https://github.com/mupq/pqm3}
\BIBentrySTDinterwordspacing

\bibitem{DBLP:journals/tches/LyubashevskyS19}
\BIBentryALTinterwordspacing
V.~Lyubashevsky and G.~Seiler, ``{NTTRU:} truly fast {NTRU} using {NTT},''
  \emph{{IACR} Trans. Cryptogr. Hardw. Embed. Syst.}, vol. 2019, no.~3, pp.
  180--201, 2019. [Online]. Available:
  \url{https://doi.org/10.13154/tches.v2019.i3.180-201}
\BIBentrySTDinterwordspacing

\bibitem{PQRISCV}
\BIBentryALTinterwordspacing
PQRISCV, ``{PQRISCV}: {Post-quantum} crypto library for the {RISC-V},''
  {Accessed}: Mar. 13, 2023. [Online]. Available:
  \url{https://github.com/mupq/pqriscv}
\BIBentrySTDinterwordspacing

\bibitem{stoffelen2019efficient}
\BIBentryALTinterwordspacing
K.~Stoffelen, ``Efficient cryptography on the {RISC-V} architecture,'' in
  \emph{Progress in Cryptology - {LATINCRYPT} 2019 - 6th International
  Conference on Cryptology and Information Security in Latin America, Santiago
  de Chile, Chile, October 2-4, 2019, Proceedings}, ser. Lecture Notes in
  Computer Science, P.~Schwabe and N.~Th{\'{e}}riault, Eds., vol. 11774.\hskip
  1em plus 0.5em minus 0.4em\relax Springer, 2019, pp. 323--340. [Online].
  Available: \url{https://doi.org/10.1007/978-3-030-30530-7\_16}
\BIBentrySTDinterwordspacing

\bibitem{adomnicai2020fixslicing}
\BIBentryALTinterwordspacing
A.~Adomnicai and T.~Peyrin, ``Fixslicing {AES}-like ciphers new bitsliced {AES}
  speed records on {ARM-Cortex} {M} and {RISC-V},'' \emph{{IACR} Trans.
  Cryptogr. Hardw. Embed. Syst.}, vol. 2021, no.~1, pp. 402--425, 2021.
  [Online]. Available: \url{https://doi.org/10.46586/tches.v2021.i1.402-425}
\BIBentrySTDinterwordspacing

\bibitem{jipeng2021}
\BIBentryALTinterwordspacing
J.~Zhang, J.~Huang, Z.~Liu, and S.~S. Roy, ``Time-memory trade-offs for
  {Saber}+ on memory-constrained {RISC-V} platform,'' \emph{{IEEE} Trans.
  Computers}, vol.~71, no.~11, pp. 2996--3007, 2022. [Online]. Available:
  \url{https://doi.org/10.1109/TC.2022.3143441}
\BIBentrySTDinterwordspacing

\end{thebibliography}

\end{document}